\documentstyle[aps,preprint,epsfig]{revtex}
\tighten
\begin{document}
\preprint{\rm FIU-NUPAR-II-\today{}}
\title {\bf Exclusive Electro-Disintegration of {\boldmath $^3$}He 
 at high {\boldmath $\boldmath Q^2$: \\ II.~Decay Function Formalism}} 
\author{M.M.~Sargsian, T.V.~Abrahamyan}
\address{Department of Physics, Florida International University, Miami, 
        Florida 33199}
\author{M.I.~Strikman}
\address{ Department of Physics, Pennsylvania State University, University Park, PA }
\author{L.L.~Frankfurt}
\address{Department of Nuclear Physics, Tel Aviv University, Tel Aviv, 
        Israel}
\maketitle
\medskip
\medskip  
\centerline{{\rm \today{}}}

\begin{abstract}
Based on the theoretical framework of generalized eikonal approximation
we study the two-nucleon emission reactions in high $Q^2$ 
electro-disintegration of $^3He$. Main aim is to 
investigate those features of the reaction which can be unambiguously identified 
with the short range properties of the ground state nuclear wave function. 
To evaluate the differential cross section we work in the formalism of the decay function
which characterizes the property of the ground state wave function as well as the decay 
properties of final two nucleon spectator system. 
Our main motivation here is to explore the accessibility of two-- and three--nucleon  short 
range correlations in $^3$He as well as to isolate unambiguously single and double 
rescattering processes in the reaction dynamics. Our analysis allowed us also to identify 
new approaches for investigating  the role of the practically unknown three-nucleon forces 
in the ground state wave function of $^3He$.

\end{abstract}

\section{Introduction}
\label{I}

In this work we study high $Q^2$ 
($4\gtrsim Q^2\gtrsim 1$~GeV$^2$) exclusive $^3He(e,e'NN)N$ 
reactions in which one nucleon in the final state can be clearly identified
as a knocked-out nucleon which carries practically all of the  momentum of 
the virtual photon.  

In the part-I of the present work\cite{SASF1}
we calculated the scattering amplitude of this type of reactions within the 
generalized eikonal approximation~(GEA), in which one 
expresses the  scattering amplitude through the sum of the diagrams
corresponding to the $n$'th order rescattering of the knocked-out  nucleon with 
the residual nucleons in the nucleus. In Ref.\cite{SASF1} we  evaluated each diagram 
based on the effective Feynman diagram rules derived within the GEA\cite{ggraphs,treview}.
The manifestly covariant nature of Feynman diagrams 
allowed us to preserve both the relativistic dynamics and the kinematics of the 
rescattering while identifying the low momentum nuclear part of the amplitude
with the nonrelativistic nuclear wave function. Such an approach allows us 
to account for the internal motion of residual target nucleons in the rescattering 
amplitude as well as finite excitation energies of residual nuclear system. 
These features of the GEA are crucial in describing  
electro-production reactions aimed at the study of short-range nuclear 
properties since these configurations are characterized by non-negligible 
values of bound nucleon momenta and excitation energies. 
The study of short-range nucleon correlations is the main goals of 
the present work. With short-range correlations we identify those interactions 
between bound nucleons that generate nucleon momenta exceeding the characteristic 
Fermi momentum of the nuclear system, $k_F\approx 250$~MeV/c. This encompasses
interactions due to the short-distance repulsive core and the short to medium-distance 
tensor interactions of NN system as well as possible three-nucleon interactions which can 
have both short and medium distance terms.

Working in the  virtual nucleon framework\cite{SASF1} one describes the 
reaction in the Lab frame relating  all non-nucleonic degrees of 
freedom effectively to the off-shellness of the knocked--out (virtual) nucleon in the nucleus.  
For recent analysis of similar reaction with two proton emission in the final state see Ref.\cite{CK}.
In calculation of the differential cross section of the reaction we utilize
the kinematics of high $Q^2$ quasielastic scattering with knocked-out nucleon 
identified  unambiguously in the final state of the reaction.
This allows us  to employ the decay function formalism in which the $eA$ cross section of 
the reaction can be represented through the convolution of the off-shell electron--bound nucleon 
cross section  and the decay function which characterizes the ground state 
properties of the nucleus as well as the decay of the spectator state in the final state 
of the reaction.

The paper is organized as follows: In Sec.~\ref{II} we summarize briefly the 
reaction, specifics of the kinematics and general form of the differential cross section.
In Sec.\ref{III} we elaborate the decay function formalism and derive the 
expression for the decay function both for plane wave  and distorted wave 
impulse approximations. For the latter case one obtains the expression based on 
GEA calculation of Ref.\cite{SASF1}.
In Sec. \ref{IV} the numerical analysis of both semi-inclusive $(e,e'N)$ and exclusive 
$(e,e'NN)$ reactions are presented. In these calculations as an input we use Bochum group's 
calculation of ground state $^3He$ wave function\cite{Bochum}, SAID group's parameterization 
of low-to-intermediate energy NN scattering amplitudes\cite{SAID} as well as our updated 
parameterization of the high energy small angle NN scattering amplitude\cite{tabra}. 
The main focus in the numerical analysis is the study of two- and three-
nucleon short range correlations~(SRCs) and isolation of the effects associated with 
single and double rescattering of the knocked-out nucleon off residual nucleons in 
the nucleus. We observe significant sensitivity of the decay function to the 
configurations  characteristic to the  short-range two- and three-nucleon correlations 
in the nucleus. The $(e,e'NN)$  reaction  provides also an ideal testing ground 
for single and double rescattering processes, which could play a crucial role 
for studies of color coherence phenomena in hard exclusive nuclear reactions.
Our calculations allowed us also to identify kinematics which could provide
unprecedented access to the three nucleon forces in $^3He$.
Section \ref{V} summarizes our results.

\section{Reaction, Kinematics and Cross Section}
\label{II}

We are studying the reaction:
\begin{equation}
e + ^3He \rightarrow e' + N_f + N_{r2}+ N_{r3}
\label{reaction}
\end{equation}
where $e$ and $e'$ are the initial and scattered electrons with four-momenta
$k_{e}$ and $k^\prime_{e}$ respectively.  The $^3$He nucleus has a four-momentum $P_{A}$.
$N_f$, $N_{r2}$ and $N_{r3}$ correspond to knocked-out and two recoil nucleons 
with four-momenta $p_f$, $p_{r2}$ and $p_{r3}$ respectively. We define also the 
four-momentum of the virtual photon $q=(q_0,{\bf q},0_\perp)\equiv k_{e}-k_{e'}$ with $Q^2=-q^2$.
The $z$ direction is chosen parallel to $\bf q$ and the scattering plane is  
the plane of the $\bf q$ and ${\bf k_{e'}}$ vectors. The missing momentum of the reaction 
is defined as: ${\bf p_m} = {\bf p_f} - {\bf  q}$.

In numerical estimations we will focus on the kinematic region:
\begin{eqnarray}
\mbox{(a)} \ 4 \ge Q^2\ge 1 \mbox{GeV}^2; \ \ \ \
\mbox{(b)} \ {\bf p_{f}}\approx {\bf q}, \ \ \ \ 
\label{kin}  
\end{eqnarray}
which allows us to identify $N_f$ as a knocked-out nucleon.  Furthermore, we will do numerical 
investigations for recoil, ${\bf p_{r2}}$, ${\bf p_{r3}}$ and missing ${\bf p_m}$ momenta 
covering the region of up to $700-800$~MeV/c (and $1~GeV/c$ for the kinematics dominated by 
the double rescattering). This exceeds the kinematic domain of 
$\le 400-500~MeV/c$, within which the  GEA and virtual nucleon approximation can be  reliably
applied\cite{SASF1}. However,  the expectation of the smooth transition to the   relativistic
region for the main properties of the decay function (see e.g. \cite{FS88}) gives some validity 
to our exploratory studies in the region of $\ge 500$~MeV/c. The properties we consider are the 
correlation relations between missing momentum/energy and momenta of the recoil nucleons, 
relative strength of 2N and 3N correlations as well as signatures identified with the 
single and double rescattering of knocked-out nucleon off spectator nucleons in the target.

\vspace{-0.4cm}
 
\subsection{Cross Section}
\label{IIb}

In numerical calculations  
we restrict ourselves by consideration of  unpolarized  three- body electro-disintegration 
reaction of Eq.(\ref{reaction}).

The differential cross section of reaction (\ref{reaction}) in which no polarizations are fixed 
is given by
\begin{eqnarray}
d^{12}\sigma&=&\frac{1}{4j_A}(2\pi)^{4}\delta^{4}(k_{e}+P_{A}-k_{e}
^{\prime}-p_{f}-p_{r2}-p_{r3})\frac{1}{4} A\cdot \sum\limits_{nucleons}
\sum\limits_{spins} |M_{fi}|^2 \nonumber\\
&&\frac{d^3 k_{e}^{\prime}}{(2 \pi)^3 2E_{e}^{\prime}} 
\frac{d^3 p_{f}}{(2 \pi)^3 2E_{f}} \frac{d^3 p_{r2}}{(2 \pi)^3 2E_{r2}} 
\frac{d^3 p_{r3}}{(2 \pi)^3 2E_{r3}} \ ,
\label{cs1}
\end{eqnarray}
where $j_A = \sqrt{(k_eP_A)^2-m_e^2M_A^2}$. 
Here we sum over final  and average over initial spins. 
The factor 1/4 comes from the averaging over the initial polarizations of 
the electron and the nucleus. 
Since one of the recoil nucleons is not observed, one eliminates this degree 
of freedom by integrating over $d^3 p_{r3}$. Thus integrated differential 
cross section is
\begin{eqnarray} 
d^{9}\sigma&=&\frac{1}{4j_A}(2\pi)^{4}
\delta(E_{e}+M_{A}-E_{e}^{\prime}-E_{f}-E_{r2}-E_{r3})\frac{1}{4} 
A\cdot \sum\limits_{nucleons}\sum_{s_{e},s_{A}}
\sum_{s_{e^{\prime}},s_{f},s_{r2},s_{r3}}|M_{fi}|^2 
\nonumber\\
&&\frac{d^3 k_{e}^{\prime}}{(2 \pi)^3 2E_{e}^{\prime}} 
\frac{d^3 p_{f}}{(2 \pi)^3 2E_{f}} \frac{d^3 p_{r2}}{(2 \pi)^3 2E_{r2}} 
\frac{1}{(2 \pi)^3 2E_{r3}} \ ,
\label{cs2}
\end{eqnarray}
where
${\bf p_{r3}}={\bf k_{e}}-{\bf k_{e}^{\prime}}-{\bf p_{f}}-{\bf p_{r2}}$.
In Eqs.\ (\ref{cs1}) and (\ref{cs2}) the transition matrix, $M_{fi}$, represents the 
convolution of the electron and nuclear currents, in which 
the nuclear current within GEA\cite{SASF1} represents the sum of the IA, single and double 
rescattering amplitudes,
\begin{equation}
M_{fi} = -4\pi\alpha {1\over q^2}j^e_{\mu}
\cdot\left(A_{0}^{\mu} + A_{1}^{\mu} + A_{2}^{\mu}\right),
\label{m}
\end{equation} 
where $A_{0}$, $A_{1}$ and $A_{2}$ are derived in Ref.\cite{SASF1}
and represent impulse approximation, single and double rescattering amplitudes respectively. 
Their final expressions are given in Eqs.(11), (21) and (27) of Ref.\cite{SASF1} respectively.
Note that IA corresponds to the impulse approximation contribution in which the 
pair distortion effects due to interaction of two outgoing recoil nucleons are taken into account.

\section{Nuclear Decay Function Formalism}
\label{III}

In description of semi-exclusive $(e,e'N)$ reactions it is conventional to introduce a 
spectral function, $S^N(E_m,{\bf p_m})$, which in the impulse approximation picture describes the probability 
of finding a struck-nucleon in the nucleus initially having  missing momentum ${\bf p_m}$ 
and missing energy $E_m= q_0 - T_f - {p_m^2\over 2(A-1)m}$. The missing energy 
characterizes the excitation energy of the final $A-1$ system.  
We generalize this approach for the case of 
the exclusive reaction in (\ref{reaction}), as well as for any semi-exclusive reactions involving 
two-nucleon emission by introducing a  nuclear decay function which can be formally defined 
as\cite{FS88}:
\begin{equation}
D^N({\bf p_m},E_m,{\bf p_{r2}}) = \sum\limits_{f}\left| \langle A-1| a^\dagger({\bf p_{r2}})a({\bf p_m})
|\Psi_{A}\rangle \right|^2\delta(E_m-(\epsilon^f_{A-2}+T_{r2}-\epsilon_A- {p_m^2\over 2(A-1)m})),
\label{dfunction}
\end{equation}
where one sums over the states, $f$,  of $A-2$ residual nucleus and 
$\epsilon^f_{A-2} = E^f_{A-2}-(A-2)m$, where  $E^f_{A-2}$ is the total energy of 
$A-2$ state.
It follows from this definition that the decay function characterizes the joint probability to 
find a struck nucleon in the nucleus  with momentum 
${\bf p_m}$, missing energy $E_{m}$ and the recoil nucleon with momentum ${\bf p_{r2}}$ 
in the decay product of the residual $A-1$ nucleus. Note that for $A\ge 4$, Eq.(\ref{dfunction}) 
assumes the sum over all possible (bound and continuum) $(A-2)$ substates, 
provided  the total excitation energy of $A-1$ final 
state is  $E_m$. The general properties of the decay function are discussed in \cite{FS88} where 
the qualitative features are studied within the two-nucleon correlation model. 
The following sum rule and normalization condition follow
from Eq.(\ref{dfunction}):
\begin{eqnarray}
& & \int d^3 p_{r2} D^N({\bf p_m},E_m,{\bf p_{r2}}) = S^N(E_m,{\bf p_m}); \nonumber \\
& & \sum\limits_{N}\int dE_m d^3p_{m} S^N(E_m,{\bf p_m}) = 1.
\label{srule}
\end{eqnarray}

Within the impulse approximation the cross section of 
the reaction with two-nucleon emission can be expressed  through the decay function as follows:
\begin{equation}
{d\sigma \over dE'_ed\Omega'_e d^3p_f d^3p_{r2}} = {j_N\over j_A}\cdot A\sum\limits_{N}
\sigma_{eN}(p_f,k_e,k'_e)\cdot D^N({\bf p_m},E_m,{\bf p_{r2}}),
\label{crsIA}
\end{equation}
where $E_m = T_{r2}+T_{r3}+|\epsilon_A| - T_{A-1}$, with $T_{A-1}$ being the 
kinetic energy of the center mass of the residual $A-1$ system.  $j_N$ is the flux 
calculated for the moving 
nucleon with momentum ${\bf p_m}$, and $\sigma_{eN}$ represents the cross section of 
electron - off-shell nucleon 
scattering\footnote{Note, that in some cases $\sigma_{eN}$ is defined without the 
flux factor (see e.g. \cite{deFor}). In these cases factor   ${j_N\over j_A}$ should not 
be included in Eq.(\ref{crsIA}).}.

Using the derivations from Sec.IIIa of Ref.\cite{SASF1} it is straightforward to calculate the decay 
function of reaction (\ref{reaction}).
According to Eq.(11) of Ref.\cite{SASF1} and Eq.(\ref{cs2}) one obtains:
\begin{eqnarray}
& & D^N({\bf p_m},E_m,{\bf p_{r2}},t_{r2}) =  {1\over 2}\sum\limits_{s_A,s_{m},s_{r2},s_{r3}}
\mbox{\huge $\vert$} \sum_{s_2,s_3}\sum\limits_{t_2,t_3}\int d^3 p_{23} 
\Psi_{NN}^{\dagger p_{r23},s_{r2},t_{r2};s_{r3},t_{r3}}(p_{23},s_2,t_2;s_3,t_3) \nonumber \\
& & \times\Psi^{s_A}_{A}(p_{m},s_m,t_m;p_{2},s_2,t_2;p_{3},s_3,t_3)\mbox{\huge $\vert$}^2 
\cdot\delta(E_m - T_{r2}-T_{r3}-|\epsilon_{A}|+{p_m^2\over 4m}).
\label{Dia}
\end{eqnarray}
Note that hereafter we use the ground state wave $\Psi^{s_A}$ normalized to A.

To understand the role of the pair distortion in the final state of the residual nucleus 
we will compare $D^N$ with  the PWIA version of the decay 
function  ($D^N_{PWIA}$), which corresponds to retaining the plane wave part of the residual 
$NN$ system's wave function, $\Psi_{NN}$ in Eq.(12) of Ref.\cite{SASF1}. 

We also can generalize the definition of the decay function to include the single and double 
rescatterings of the knock-out nucleon with residual nucleons. Such a generalization usually is 
meaningful within an approximation in which electromagnetic current of knocked-out nucleon
 is factorized from rescattering integrals of Eqs.(21) and (27) of Ref.\cite{SASF1}.  
Two conditions should be met in order for this factorization  to be justified:
\begin{itemize}
\item First, one should be able to neglect the charge-exchange part of the amplitude for
high energy small angle $NN$ rescatterings. This  follows, first, 
from the fact that at sufficiently high energies the energy dependence of the 
small angle hadronic scattering amplitude scales as $\sim s_{NN}^{J-1}$, 
where $J$ corresponds to the spin of the exchanged particle \cite{Azimov}.
Since, at energies  considered,  charge-exchange requires predominantly spin-0 exchange  
in the $t$-channel as compared to spin-one exchange in the diagonal channel, one expects 
strong $s$ suppression of  the charge-exchange amplitude as compared to the elastic $NN$ amplitude.
Analysis of the existing data on NN scattering demonstrates (see e.g. \cite{GibbsLo})
that starting at $Q^2\ge 2$ GeV$^2$ the charge-exchange part amounts only a third of 
the forward scattering $NN$  amplitude and decreases linearly with an increase of $Q^2$.  
Moreover, this contribution is further suppressed due to the fact that 
the charge-exchange amplitude is predominantly real and does not interfere with the dominant 
part of the FSI amplitudes which are predominantly imaginary. 
Taking into account also the larger slope factor of the $t$ dependence of charge-exchange 
amplitude as compared to the forward scattering, one estimates only a few percent overall 
contribution  to the $NN$ rescattering amplitude at $Q^2\ge 2$ GeV$^2$. This contribution 
decreases sharply with an increase of $Q^2$.


\item Secondly, the transfered momentum, $q$,   should be large enough that
one can neglect the $k$, $k_2$ and $k_3$ dependencies in the electromagnetic current in 
Eqs.(21) and (27) of Ref.\cite{SASF1}. Such an approximation is justified by the kinematic 
condition of Eq.(\ref{kin}) and by the fact that  the characteristic average momenta 
transferred during $NN$ rescattering are restricted by the slope of the exponent 
in Eq.(A2) of Ref.\cite{SASF1}, i.e. $\langle k^2\rangle, \langle k_2^2\rangle, 
\langle k_3^2\rangle \le  {2\over B}\sim 250$~(MeV/c)$^2$. 
\end{itemize}

Keeping only the diagonal part in the $NN$ rescattering amplitude and factorizing the 
electromagnetic current of knocked-out nucleon, we arrive at an expression similar to 
Eq.(\ref{crsIA}) in which we  refer $D$ as a decay function calculated within  
the distorted wave impulse approximation~(DWIA), $D_{DWIA}$.  
Based on Eqs.(11,21,27) of Ref.\cite{SASF1}, for $D_{DWIA}$ one 
obtains:
\begin{eqnarray}
& & D^N_{DWIA}(Q^2, {\bf p_m},E_m,{\bf p_{r2}},t_{r2}) = {1\over 2}\sum\limits_{s_A,s_{m},s_{r2},s_{r3}}
\nonumber \\
& & \mbox{\huge $\vert$}
\sum\limits_{s_2,s_3}\sum\limits_{t_2,t_3}\left\{ \int d^3 p_{23} 
\Psi_{NN}^{\dagger p_{r23},s_{r2},t_{r2};s_{r3},t_{r3}}(p_{23},s_2,t_2;s_3,t_3) 
\Psi^{s_A}_{A}(p_{m},s_m,t_f;p_{2},s_2,t_2;p_{3},s_3,t_3) \right.  \nonumber \\
& & -{1\over 2} \int {d^3 k d^3 p_{23}  \over (2 \pi)^3} 
\Psi_{NN}^{\dagger p_{r23},s_{r2},t_{r2};s_{r3},t_{r3}}(p_{23},s_{2},t_2;s_3,t_3) 
{1\over \Delta^0-k_z+i\varepsilon}   \nonumber \\
& & \times \left[ \chi_1(s^{NN}_{2})f^{t_f,t_{2}|t_f,t_2}_{NN}(k_{\perp})
\cdot \Psi^{s_A}_{A}(p_{m}+k,s_m,t_f;-{p_m\over 2} + p_{23}-k,s_2,t_2;
 -{p_m\over 2} - p_{23},s_3,t_3) \right. \nonumber \\ 
& & \left. \ \  +  \chi_1(s^{NN}_{3})f^{t_f,t_{3}|t_f,t_3}_{NN}(k_{\perp})\cdot \Psi^{s_A}_{A}(p_{m}+k,s_m,t_f;-{p_m\over 2} + p_{23},s_2,t_2;
 -{p_m\over 2} - p_{23}-k,s_3,t_3) \right]  \nonumber \\
& & + {1\over 4}\int d^3 p'_{23}  {d^3k_3\over (2\pi)^3}{d^3 k_2 \over (2 \pi)^3} 
\Psi_{NN}^{\dagger p_{r23},s_{r2},t_{r2};s_{r3},t_{r3}}(p'_{23},s_{2},t_2;s_{3},t_3) 
\cdot f^{t_3,t_{f}|t_3,t_f}_{NN}(k_{3\perp})f^{t_2,t_{f}|t_2,t_f}_{NN}(k_{2\perp})\nonumber \\ 
& & \times \left[  
{\chi_2(s^{NN}_{b3})\over \Delta_3 - k_{3z} + i\varepsilon} 
{\chi_1(s^{NN}_{a2})\over \Delta^0 -k_{2z}-k_{3z}+i\varepsilon} + 
{\chi_2(s^{NN}_{b2})\over \Delta_2 - k_{2z} + i\varepsilon} 
{\chi_1(s^{NN}_{a3})\over \Delta^0 -k_{2z}-k_{3z}+i\varepsilon} \right]
\nonumber \\
& & \times \left. \Psi^{s_A}_{A}(p_{m}+k_3+k_2,s_m,t_f;-{p_m\over 2}-k_2+p'_{23},s_2,t_2;
-{p_m\over 2}-k_3 - p'_{23},s_3,t_3) \right \}
\mbox{\huge $\vert$} ^2 \nonumber \\
& & \times\delta(E_m - T_{r2}-T_{r3}-|\epsilon_{A}|+{p_m^2\over 4m}).
\label{Ddwia}
\end{eqnarray}
Further in the text we will refer $D_{DWIA}$ also as $D_{IA+FSI}$
\footnote{The cases in which only a single rescattering is considered we refer to as
IA+FSI1.}.

\section{Numerical Results}
\label{IV}
In our numerical analysis we use the following inputs:
\begin{itemize}
\item {\bf Ground State Wave function:} We use the $^3He$ ground state wave functions calculated 
by the Bochum group\cite{Bochum} solving Faddeev equation for different sets of realistic 
$NN$ potentials. Authors in Ref.\cite{Bochum} consider also the different models 
of 3N forces with the main motivation to describe the binding energy of the $A=3$ system.
\item {\bf Pair Distortion:} To estimate the reinteraction of two outgoing slow nucleons 
in Eq.(12) of Ref.\cite{SASF1} we use the SAID group's parameterization of the NN scattering 
amplitudes based on the partial wave analysis of the $NN$ scattering data\cite{SAID}. 
This parameterization 
successfully describes the $NN$ scattering data at a low to intermediate momentum range 
($p_{LAB}\le 1.3/3$~GeV/c for $pn/pp$ scattering).   
\item{\bf FSI:} To estimate the single and double rescattering contributions in the total
scattering amplitude, we use the following conventional parameterization for high energy 
small angle $NN$ scattering:
\begin{equation}
f^{NN} = \sigma_{tot}^{NN}(s)(i+\alpha^{NN}(s))e^{{B^{NN}(s)\over 2}t}\delta^{h_1,h^\prime_1}
\delta^{h_2,h^\prime_2},
\label{fnnp}
\end{equation}
where we parameterized\cite{tabra} $\sigma_{tot}^{NN}$,  $\alpha^{NN}(s)$ and $B^{NN}(s)$ 
for both $pn$ and $pp$ scattering. We use practically all published data on small angle 
nucleon-nucleon scattering.

In performing the numerical integrations in Eqs.(21) and (27) of Ref.\cite{SASF1} we  
represent  the knocked-out nucleon propagator in the integral as a  sum of the pole 
and principal value~(P.V.) terms 
 \begin{equation}
{1\over \Delta-k_{z}+i\epsilon}=-i\pi \delta(\Delta-k_{z})  + P.V. {1\over \Delta - k_z},
\label{polandPV}
\end{equation}
where $k_z$ and $\Delta$  characterize the transfered longitudinal momenta 
due to  rescattering. Note that  strictly speaking one can use the parameterization of 
the free (on-shell) NN scattering amplitude in Eq.(\ref{fnnp}) for the pole term of  
Eq.(\ref{polandPV}) only. For the P.V. term, the  NN rescattering amplitude is off-shell. 
However, we use the parameterization in Eq.(\ref{fnnp}) for P.V. term too since, in the 
kinematics of interest, the P.V. term is only a small correction as compared to the  
pole contribution of Eq.(\ref{polandPV}).

\end{itemize}

\medskip
\medskip

In the following numerical analyzes our main goal is to  identify the strategies which are 
best suited for investigation of two- and three- nucleon short range correlations. 

\begin{figure}[ht]
\centerline{\epsfig{height=8cm,width=12cm,file=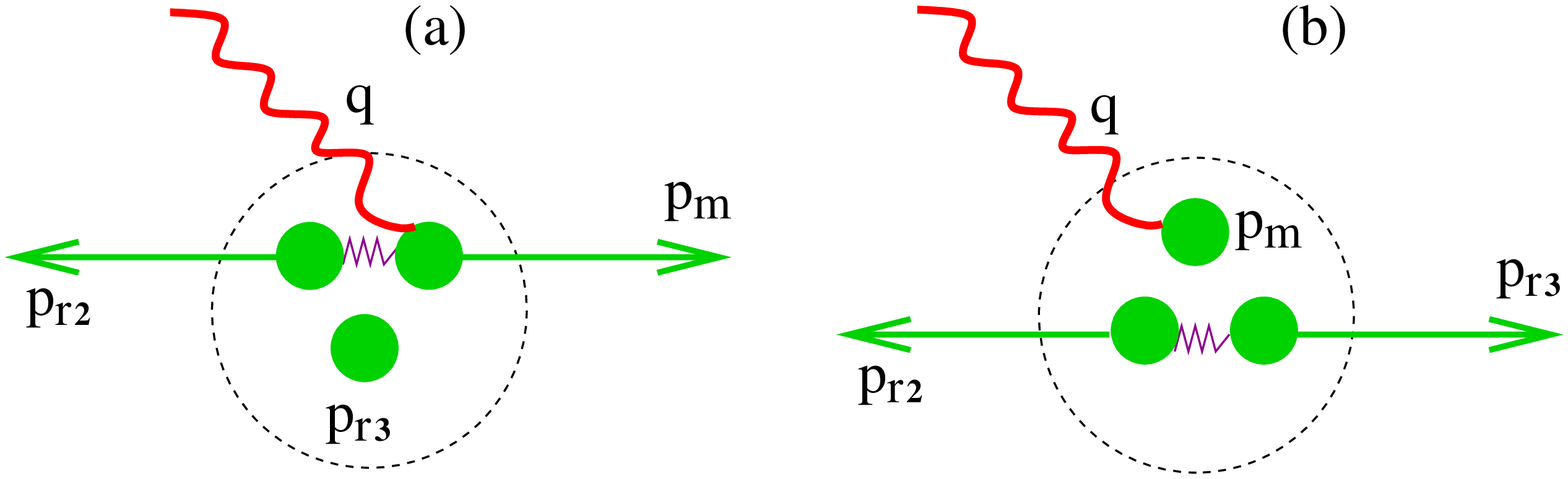}}
\caption{(Color online) Definition of type 2N-I (a) and type 2N-II (b) correlations.}
\label{2Nsrc}
\end{figure}

For {\bf two nucleon} correlations 
we are particularly interested in isolation and studies of the following configurations: 
\begin{itemize}
\item {\bf Type 2N-I} correlations from which a nucleon is knocked-out by virtual photon, 
while the third nucleon moves in the mean field of 2N SRC Fig.\ref{2Nsrc}(a). 
For the idealized case of third nucleon at rest one obtains a correlation relation:
\begin{equation}
E_{m}^{(2N-I)} = \sqrt{m^2+p_m^2} - m - T_{A-1},
\label{empm_2N-I}
\end{equation}
where $T_{A-1}=\sqrt{4m^2+p_m^2}-2m$. Note that although we apply the approximation, $p^2/m^2\ll 1$ 
in calculation of the scattering amplitudes, we still use the relativistic form of the kinetic 
energy to preserve the relativistic kinematics.

\item {\bf Type 2N-II} correlations in which virtual photon strikes a third isolated nucleon 
while 2N SRC breaks up at the final state of the reaction, Fig.\ref{2Nsrc}(b). 
In this case the expected correlation between $E_m$ and $\bf {p_{r2}}\approx - {\bf p_{r3}}$ 
for the case of  $p_m\approx 0$ is:
\begin{equation}
E_{m}^{(2N-II)}  =  \sqrt{m^2+p_{r2}^2} +  \sqrt{m^2+p_{r3}^2} - 2m
\label{empm_2N-II}
\end{equation}
\end{itemize}

\begin{figure}[ht]
\centerline{\epsfig{height=8cm,width=12cm,file=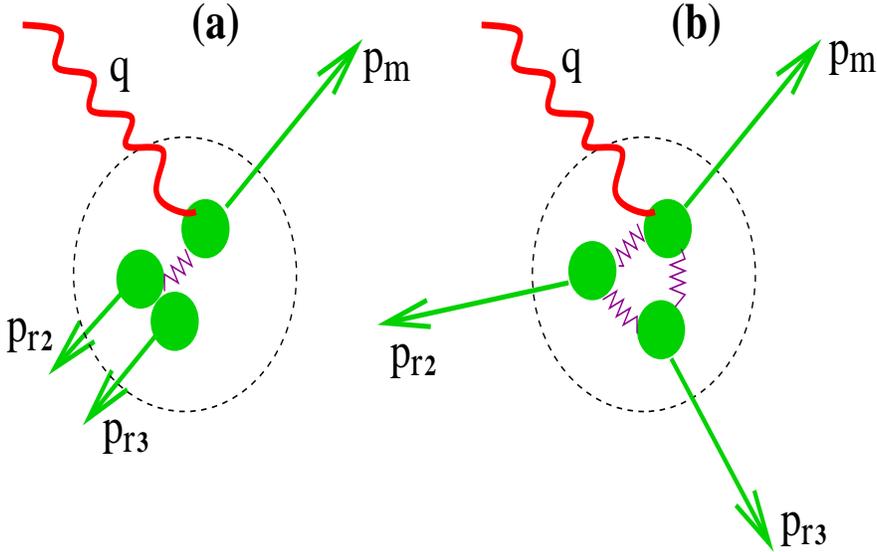}}
\caption{(Color online) Definition of type 3N-I~(a) and type 3N-II~(b) correlations.}
\label{3Nsrc}
\end{figure}

For {\bf three nucleon} correlations we focus on two particular situations:
\begin{itemize}
\item {\bf Type 3N-I} correlations, in  which  case the high missing 
momentum of the knocked-out nucleon is balanced by two nucleons which share almost equally the 
missing momentum $p_m$, Fig.\ref{3Nsrc}(a). This corresponds to the minimal missing mass of 
recoil 2N system with missing energy:
\begin{equation}
E_m^{(2N-I)} \approx |\epsilon_A|.
\label{empm_3N-I}
\end{equation}
\item {\bf Type 3N-II} correlations, in  which case the nucleon is knocked-out from the 
symmetric configurations where all three nucleons have the same  momentum 
$p_m$, Fig.\ref{3Nsrc}(b). This corresponds to a significantly higher value of $E_m$ as compared to the 
the above discussed 2N and 3N SRC cases:
\begin{equation}
E_{m}^{(3N-II)}  =  2\sqrt{m^2+p_{m}^2} - 2m - T_{A-1}.
\label{empm_3N-II}
\end{equation}
\end{itemize}

\begin{figure}[ht]
\vspace{-0.2cm}
\centerline{\epsfig{height=10cm,width=10cm,file=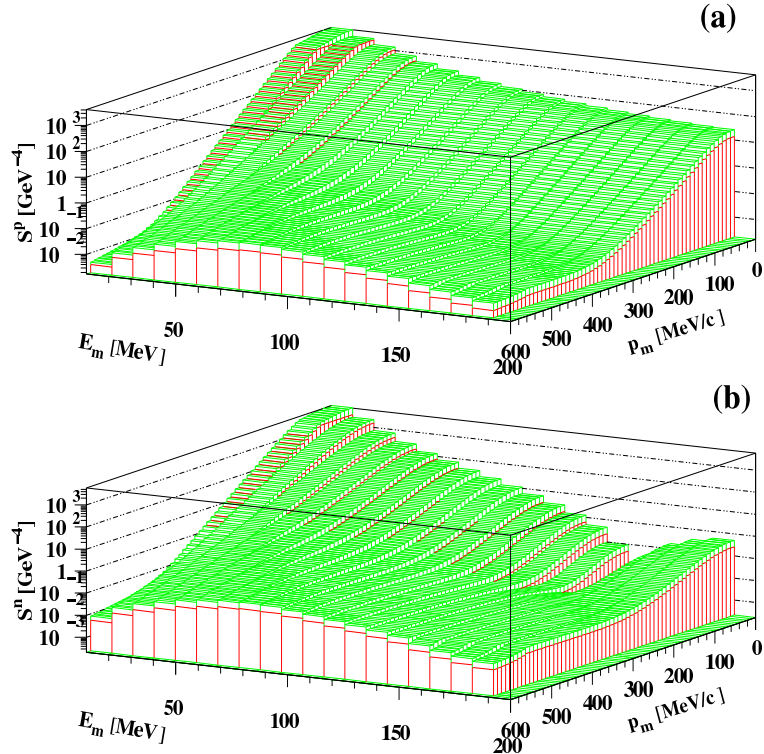}}
\caption{(Color online) Spectral function for $^3He(e,e'N)X$ reaction for struck proton~(a) and 
neutron~(b) calculated within the PWIA and IA. The lines at the sides of the 
histograms shows the effects due to pair distortion. }
\label{spectral}
\end{figure}

\subsection{Spectral Function}
We start with the calculation of  the spectral function as it is defined in Eq.(\ref{srule}).
This quantity is relevant primarily for semi-exclusive $(e,e'N)$ reactions, in which 
only the scattered electron and knocked-out nucleon are registered. 
Figure \ref{spectral} shows the spectral functions for 
the cases of knocked-out proton~(a) and knocked-out neutron~(b)  
calculated with the $^3He$ ground state wave function based on 
the \mbox{Urbana-IX} potential\cite{Bochum,Urbana}. 
The hatched surface represents the PWIA prediction  while dotted contours show the 
effect of pair distortion.  

The correlation feature of Eq.(\ref{empm_2N-I}) is reflected in 
the emergence of the broad peak in $E_m$ distribution at $p_m\ge 300$~MeV/c, while the signature of 
2N-II correlations is seen in the minimum of the $E_m$ distribution at $p_m\approx 0$ for the 
case of knocked-out neutron with two recoil protons. This 
minimum reflects the fact that the relative momentum distribution 
in the $pp$ pair has a node in the $S$-state at $\approx 420$~MeV/c.
These results are in agreement with  previous  analyses of spectral function\cite{Ciofi}.

\begin{figure}[th]
\vspace{-0.8cm}
\centerline{\epsfig{height=12cm,width=12cm,file=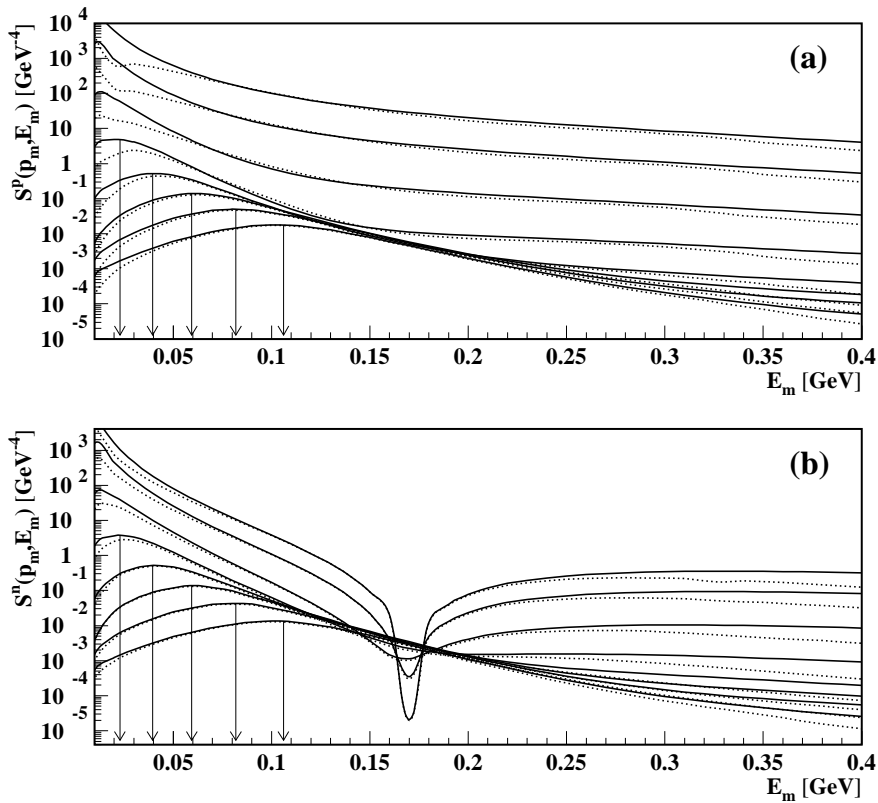}}
\caption{Missing energy distribution of the spectral function for the removal of 
the proton~(a) and neutron~(b) at different values of 
missing momenta. Curves counted from the above at $E_m=0$ correspond to the missing momenta 
$0$ to $700$~MeV/c with $100$~MeV/c increments. Solid and dotted lines correspond to 
PWIA and IA predictions respectively. Arrows correspond to the prediction of Eq.(\ref{empm_2N-I})-
scattering off a quasi-free stationary two-nucleon correlation.}
\label{spectralsrc}
\end{figure}

The emergence of these correlations at $p_m\ge 300$~GeV/c are clearly seen 
in Fig.(\ref{spectralsrc}), in which 
solid and doted lines represent calculations within PWIA and IA approximations 
respectively.
The positions of the peaks can be related to the dominance of type  
2N-I SRCs in the high missing momentum part of the nuclear 
wave function \cite{FS81,CSFS91} according to Eq.(\ref{empm_2N-I}) (i.e. $E_{m}^{Peak}\approx
E_{m}^{2N-I}$).  The solid arrows in the figure correspond to the prediction of 
Eq.(\ref{empm_2N-I}) corresponding to the 
scattering off a quasi-free and stationary two-nucleon correlation. 
This situation is reflected in the fact that, at large nucleon momenta $p_m$, the 
spectral function has similar functional dependence on $p_m$ as the stationary two nucleon 
correlation \cite{FS81}. It is worth noting that there were several experimental 
indications\cite{Saclay,NIKHEF} of $E_m-p_m$ correlations according to Eq.(\ref{empm_2N-I}). 
However, the lower values of $Q^2$ in these experiments did not allow direct discrimination 
of the short-range two-body forces from the long-range two-body currents corresponding to the  
meson-exchange contributions.

The type 2N-II SRC can be identified clearly only for the spectral function involving two 
protons in the recoil kinematics, Fig.\ref{spectralsrc}(b).  As Figs.\ref{spectral} and 
\ref{spectralsrc} demonstrate, no 
clear signatures are seen for 3N correlations. One expects type 3N-I correlations to dominate 
at the left corner of the $E_m$ distribution starting at $p_m\ge 400-500$~MeV/c, while type 
3N-II correlations dominate at the right (higher value)  corner of the $E_m$ distribution 
starting at $p_m\gtrsim 400$~MeV/c. Fig.\ref{spectralsrc} also demonstrates a strong dependence of  
pair-distortion effects on considered  values of  $E_m$ and $p_m$.

\subsubsection{Pair Distortion Effects}

\begin{figure}[t]
\vspace{-0.8cm}
\centerline{\epsfig{height=11cm,width=11cm,file=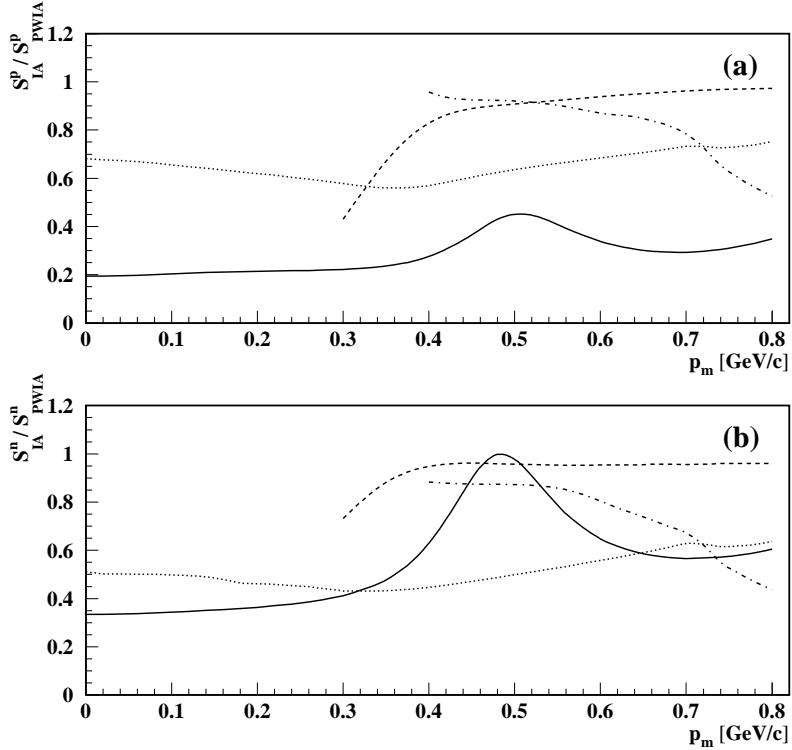}}
\caption{Missing momentum dependence of the ratio of spectral function calculated
within the IA and  PWIA approximation. Solid lines correspond to  $E_m=0$. For the 
dashed lines, $E_m$ is defined according to Eq.(\ref{empm_2N-I}). 
For the dashed-dotted lines, $E_m$ defined 
according to Eq.(\ref{empm_3N-II}) and for dotted lines, $E_m=350$~MeV.}
\label{spectralpd}
\end{figure}

The pair distortion effects can be assessed quantitatively 
in Fig.\ref{spectralpd}  which shows the missing momentum, $p_m$,  
distribution of the ratios of the spectral functions calculated within the PWIA and IA 
approximations
for different values of $E_m$. These calculations demonstrate that the 
pair distortion strongly suppresses  (by a factor of five) 
the spectral function at the  kinematic region of small $p_m$ and $E_m$. 
This reflects the fact that in this region two spectator 
nucleons have vanishing relative momentum  at which the interaction cross section is very large.  
For the same reason, one observes  significant pair distortion effects in the kinematics 
(see Eq.(\ref{empm_3N-I})) favorable for studies of type 3N-I correlations 
(high $p_m$ part of the solid curve in Fig.(\ref{spectralpd})). Note that in this case 
pair distortion is larger for the $pn$ recoil case than for the $pp$ case.

Pair distortion is also large in the kinematics  of very  large $E_m > E_m^{2N-I,II}$ 
(dotted and dash doted lines in Fig.\ref{spectralpd})  where one expects the dominance of 
type 3N-II correlations. The large pair distortion effects in the kinematics of 
3N correlations can be understood qualitatively. The pair distortion effectively represents
the three-nucleon correlation, in which the initial short range $NN$ 
correlation between knocked-out nucleon and one of the spectator nucleons is combined with 
the final state $NN$ reinteraction between two recoil nucleons.

For type 2N-I correlations, Fig.\ref{spectralpd} (dashed lines) reveals modest pair distortion 
effects starting at $p_m\ge 400$MeV/c. At the same time,  Fig.\ref{spectralpd} demonstrates more 
pair distortion for type 2N-II  correlations. This result can also be understood qualitatively.
For type 2N-I correlations, one of the recoil nucleons is initially in the SRC, while the 
second one  is separated from the SRC with  relatively small momentum. 
As a result, once one of the nucleons is instantaneously removed 
from the SRC, the two recoil nucleons in the final state will be spatially separated, thus 
reducing the probability for their interaction.

\begin{figure}[ht]
\vspace{-0.8cm}
\centerline{\epsfig{height=11cm,width=11cm,file=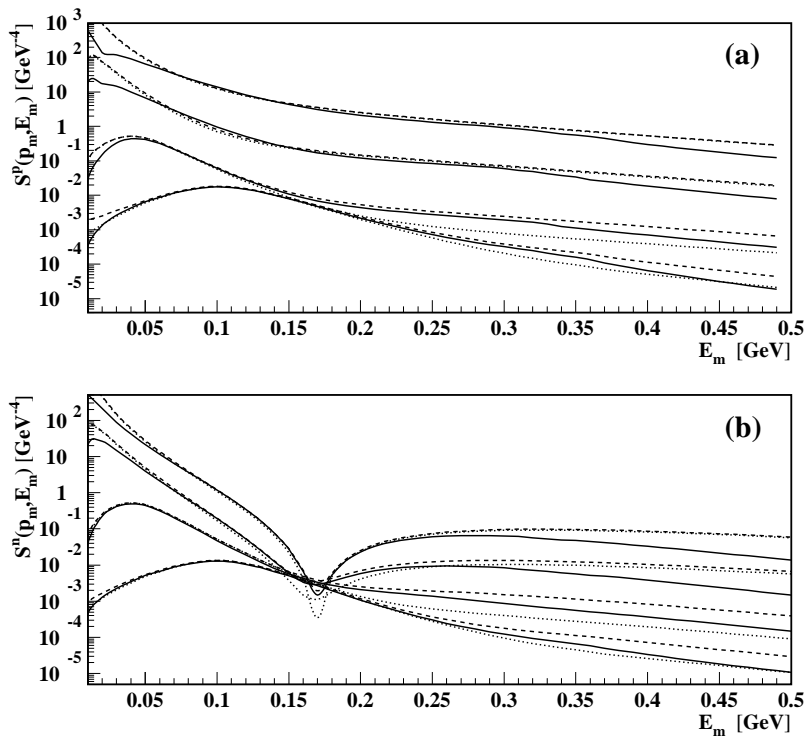}}
\caption{Missing momentum dependence of the spectral function at different values of 
missing momenta. The curves counted from the upper left to the lower left of the 
figures correspond to the missing momenta $100$, $200$, $400$ and $700$~MeV/c. 
Dotted lines -- PWIA prediction with only NN  potential considered, dashed line -- 
PWIA with additional 3N forces included, solid line -- IA with the same 3N forces included.}
\label{spectral3m}
\end{figure}

The situation is different for type 2N-II correlations in which two 
recoil nucleons are in the SRC and are spatially close to each other.
This will result in much larger pair distortion effects than for type 2N-I correlations.
The observed feature is consistent with the qualitative arguments of 
Ref.\cite{FS81,FS88} that to break the SRC and release a  spectator from 
it  with minimal distortion it is preferable to knock-out a nucleon directly from the SRC.

The next question we address is whether the spectral function can provide an effective framework 
for studies of the implication of three nucleon forces~(3NFs) in the ground state wave 
function of $^3He$. Qualitatively, one expects that the onset of 3NFs to 
arise in the kinematics dominated by three-nucleon correlations.
This expectation is confirmed in Fig.\ref{spectral3m} where the main 3NF effects are seen at 
$E_m=|\epsilon_A|$ and $E_m>  E_{m}^{2N-I,II}$. Here we consider the 
difference in the predictions of the spectral functions calculated based on the 
$^3He$ wave functions which are  calculated with and without explicit inclusion of 3NFs 
(see e.g. \cite{Bochum}).  Predictions are made 
within PWIA are denoted by dotted and dashed lines in Fig.\ref{spectral3m}.

However, as it was discussed above, in the same region of $E_m$ and $p_m$ we predict sizable 
effects due to  pair distortion, which effectively imitates a three-nucleon correlation. 
As Fig.\ref{spectral3m} demonstrates pair distortion~(solid lines)  will considerably 
diminish signatures related to 3NF effects in the spectral function.

\begin{figure}[ht]
\vspace{-0.8cm}
\centerline{\epsfig{height=11cm,width=11cm,file=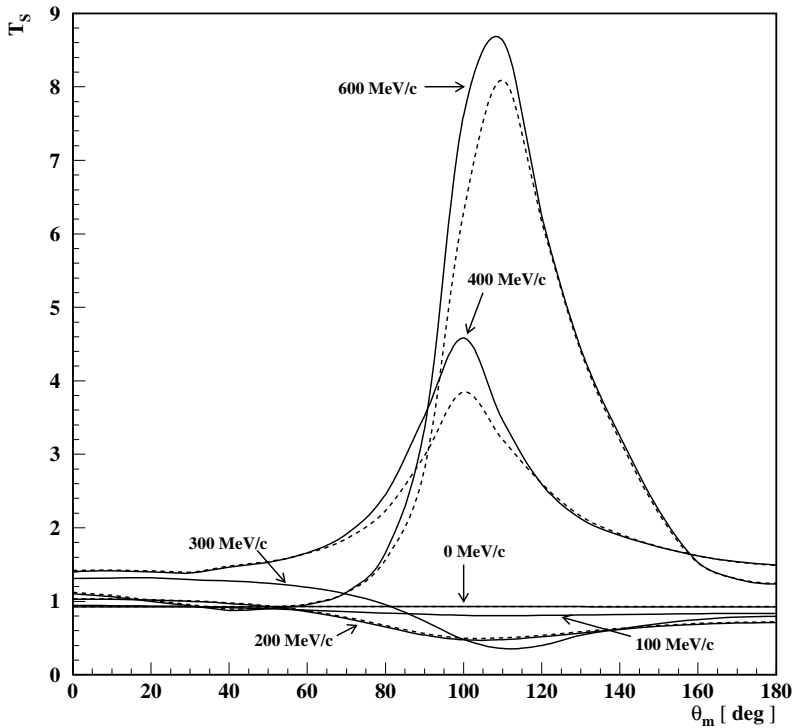}}
\caption{Angular dependence of transparency $T$ as defined in Eq.(\ref{trans}) for the reaction 
with struck proton at different values of missing momenta. The missing energy here is 
defined according to type 2N-I correlation condition of Eq.(\ref{empm_2N-I}). 
Solid line -- only single rescattering amplitude is included in FSI, dashed line -- 
both single and double rescattering amplitudes are included in FSI.}
\label{FSI_th}
\end{figure}

\subsubsection{Final state interaction~(FSI) effects}
\label{Sec.Spec.FSI}

The inclusion of  the FSI of the knocked-out nucleon with spectator nucleons removes 
the isotropy of spectral function with  respect to the direction of the virtual photon, ${\bf q}$.
To asses the FSI effects quantitatively, we analyze the ratio, $T_S$, defined as follows:
\begin{equation}
T_S = {S^N_{DWIA}(Q^2,q,E_m,p_m)\over S^N_{IA}(E_m,p_m)},
\label{trans}
\end{equation}
where subscripts $DWIA$ and $IA$ denote the spectral functions calculated with and without 
FSI effects. Here $T_S\approx 1$ corresponds to the small 
effects of the FSI, while $T>1$ or $T<1$ will 
indicate the dominance of the absorption or the rescattering effects due to the FSI.
Based on the analysis of the analytic properties of rescattering amplitudes in GEA~\cite{ggraphs}, 
it is possible to identify the kinematic regions in which one can isolate the FSI or the SRC  
as the dominant term in the scattering amplitude.

Similar to the reaction of exclusive electro-disintegration of the deuteron\cite{deuteron}, one 
expects the  FSI  to dominate  in  nearly transverse kinematics in which  $\alpha_m=1$ and 
$p_{m}\ge 100$~MeV/c. 
Here, $\alpha_m$ in the IA describes the momentum fraction of the nucleus carried out by a 
nucleon ``1'' in the infinite momentum frame of the nucleus. It is defined  as follows:
\begin{equation}
\alpha_m = {p_{m-}\over p_{A-}} \approx {p_{f-} - q_{-}\over m}\mid_{Lab \ Frame} ,
\label{lcom}
\end{equation}
where the ``minus" components of momenta are defined by $k_- = k_0-k_z$.

Fig.\ref{FSI_th} represents the $\theta_m~(\alpha_m)$ dependence of $T_S$ at different values of 
$p_m$ calculated at $Q^2=4$~GeV$^2$. The missing energy in these calculations is chosen
for the type 2N-I correlation (Eq.(\ref{empm_2N-I})). The figure demonstrates 
 the presence of  significant FSI effects in the  near-transverse kinematics, $\alpha_m\approx 1$, with the effects diminishing 
at parallel~($\theta_m=0^0$)  and anti-parallel~($\theta_m=180^0$) kinematics
\footnote{Everywhere in the text  
parallel/anti-parallel kinematics correspond to $\theta_m=0^0/180^0$}.

The missing energy $E_m$, gives us an additional degrees of freedom to ``manipulate'' 
the relative strength of the FSI and SRC  effects in the different kinematics of 
electro-disintegration.  
This is especially important for isolating SRC structures in the spectral function as it 
is measured in $A(e,e'N)X$ reaction. In the beginning of Sec.\ref{IV}, we identified 
several kinematic regions in which the strength of the spectral function is largely  determined by 
SRCs. In all these cases, the initial momenta of at least two nucleons in the nucleus exceeds 
$300-400$~MeV/c. The final state reinteraction of the knocked-out nucleon with the spectator 
nucleons can largely destroy this picture. For example, as can be seen from 
Eqs.(21) and (27) in Ref.\cite{SASF1}, due to the integration 
in the rescattering loops, it is impossible to ensure the appearance of large 
values of internal momenta in the ground state wave function of 
the nucleus in a straightforward way.
Thus,  the condition  $p_m\ge 300$~MeV/c or $E_m\gtrsim 100$~MeV  may not ensure the dominance of 
the high-momentum component in the ground state wave function of the nucleus. 
This situation may significantly affect the identification of type 2N and 3N correlations.

The problem of suppression of the FSI in probing the SRC in the $A(e,e'N)X$ reaction was addressed 
within the GEA in Ref.\cite{ggraphs}. It was observed that a trivial condition, $p_m\ge k_F$, is not 
sufficient to probe the SRC  component of the ground state nuclear wave function. 
One needs to impose the following additional condition on the  effective longitudinal 
momentum, $p_Z$, entering  the single rescattering amplitude,
\begin{equation}
p_Z \equiv p_{mz} + {q_0\over q}(E_m + {p_m^2\over 4m}) > k_F.
\label{pZ}
\end{equation}
With this, one will ensure that rescatterings happen within the SRC. Note that ``$p_Z$''s
in Eq.(\ref{pZ}) correspond to the pole values of fast  nucleon propagators  in the 
rescattering amplitude (see e.g. Eq.(21) in Ref.\cite{SASF1}).

\begin{figure}[ht]
\vspace{-0.8cm}
\centerline{\epsfig{height=12cm,width=12cm,file=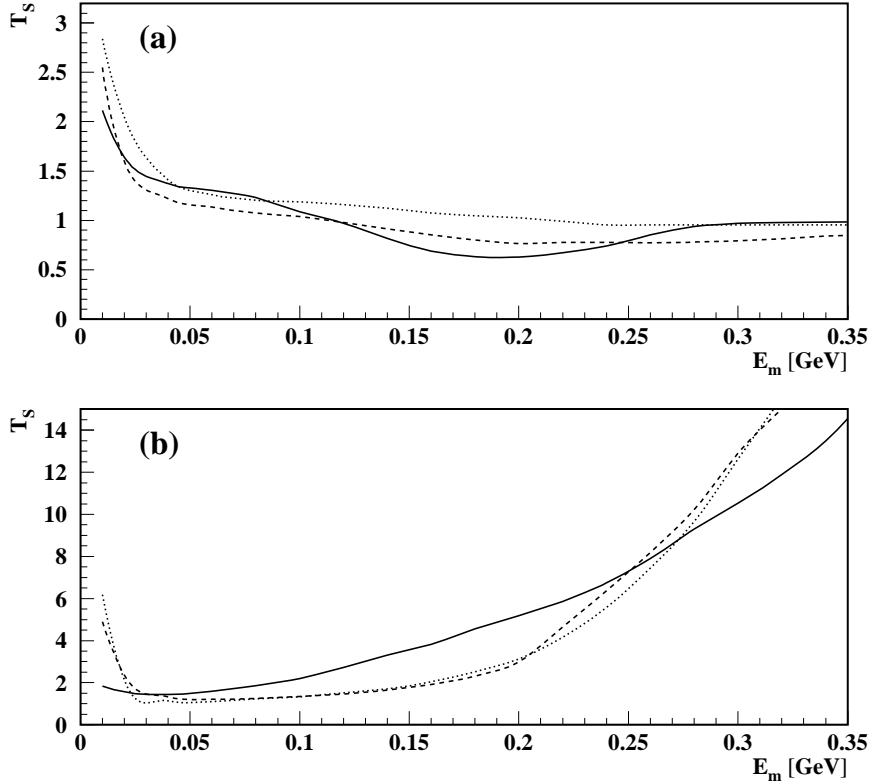}}
\caption{Missing energy dependence of transparencies for the knocked-out proton reaction at 
parallel, $\theta_m=0^0$, (a)  and anti-parallel, $\theta_m=180^0$, (b) kinematics. 
The solid, dashed and dotted curves corresponds to the calculation with missing momenta 
equal to $400$, $500$ and $600$~MeV/c respectively.}
\label{fig_pZ}
\end{figure}

In Fig.\ref{fig_pZ}, we consider the $E_m$ ($p_Z$) dependence of $T_S$  for 
parallel~(a) and anti-parallel~(b) kinematics 
for values of $p_m$ equal to $400$, $500$ and $600$~MeV/c. In Fig.\ref{fig_pZ}(a), 
the center of mass momentum  of the two recoil nucleons is in the direction backward to 
${\bf q}$, while in Fig.\ref{fig_pZ}(b), the recoil system is produced in the forward direction. 
One can see from  this figure that the FSI indeed decreases with increase of $|p_Z|$. 
This indicates that the FSI is increasingly confined within the SRC. An interesting feature 
of these results is that while the FSI contribution keeps decreasing with increase of $|p_Z|$ 
in the kinematic region relevant for type 2N-I and 3N-II correlations, it grows sharply in 
the region associated with the type 3N-I correlations. 
Both trends can be understood qualitatively. When $|p_Z|\le k_F$, one has dominant 
contributions from the FSI involving uncorrelated nucleons which have a larger probability 
amplitude in the ground state wave function of $^3He$.
Once $|p_Z|> k_F$, the FSI takes place predominantly within the
2N correlation. In the latter case, for type 2N-I SRC, the third nucleon has a small 
momentum and is spatially separated from the NN SRC. Thus, it does not contribute substantially 
to the FSI. 

\begin{figure}[ht]
\vspace{-0.8cm}
\centerline{\epsfig{height=12cm,width=12cm,file=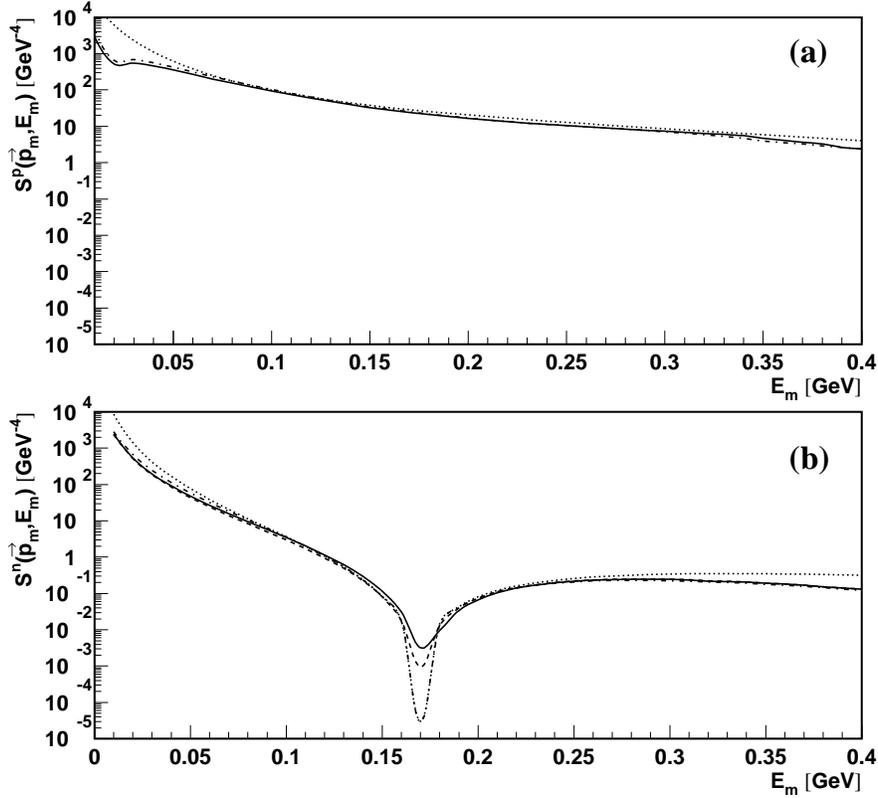}}
\caption{Missing energy dependence of the spectral function for the type 2N-II correlation 
condition at missing momentum $p_m=0$. Dotted, dashed-dotted, dashed and solid lines correspond to 
PWIA, IA, IA+FSI1 and full~(IA+FSI) calculations respectively.}
\label{FSIem_pn_2N-II}
\end{figure}

For type 3N-I correlations both spectator nucleons contribute to the FSI 
since they are both correlated with the knocked-out nucleon.
This results in larger  FSI effects as compared 
to the IA contribution. These calculations indicate that 
(except for the case of type 3N-II SRC in the parallel kinematics) the FSI generally dominates 
in the kinematics where one expects an enhanced contribution from 3N SRCs in the 
wave function of $^3He$.

For type 2N-II correlations, the FSI takes place between the 
knocked-out nucleon, which is initially almost at rest and spatially separated from the 2N SRC 
(see Fig.\ref{2Nsrc}(b)). As a result, one expects a diminished FSI  contribution in 
whole range of $E_m$. Such behavior can be seen in Fig.\ref{FSIem_pn_2N-II} 
for the calculated  $E_m$ dependence of the spectral function 
at  $p_m=0$. This calculation shows that the type 2N-II SRC for the knocked-out neutron 
reaction attains the characteristic minimum in  $E_m$ distribution although  
smeared out strongly due to rescatterings.

\begin{figure}[ht]
\vspace{-0.8cm}
\centerline{\epsfig{height=12cm,width=12cm,file=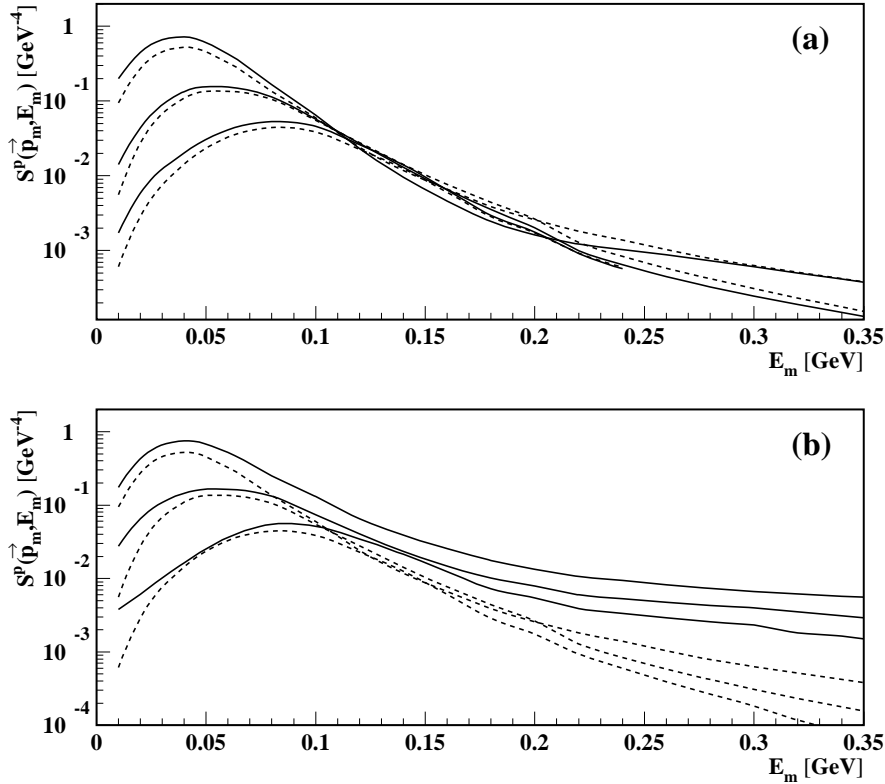}}
\caption{Missing energy dependence of the spectral function for the knocked-out proton reaction in 
parallel~(a) and anti-parallel~(b) kinematics at different values of missing momenta.
The upper, middle and lower curves at  the left corner of the figure correspond to the 
missing momenta $400$, $500$ and $600$~MeV/c respectively. Dashed line -- IA prediction, 
solid line IA+FSI prediction.}
\label{FSIem}
\end{figure}

In Fig.\ref{FSIem}  we summarize the predictions for the $E_m$ dependence 
for the spectral function at different values of missing momenta $p_m$ calculated for 
parallel (a) and anti-parallel (b)  kinematics. 
These calculations demonstrate that the observed $E_m-p_m$ correlation  within the IA  
for type 2N-I SRC  generally survives the FSI for both parallel and anti-parallel kinematics. 
For  3N correlations, type 3N-II SRC survives FSI for parallel kinematics while 
type 3N-I SRC is strongly affected  by FSI.

Summarizing our consideration of the spectral function we can conclude that the $A(e,e'N)X$ 
reaction is best suited for studies of 2N correlations only. Type 2N-I SRC survives both pair 
distortion and the FSI while type 2N-II attains its specific feature for neutron-knock out 
reactions at $p_m\le 100$~MeV/c \footnote{As a note of caution, we point out that the 
minimum in the $E_m$ distribution is very narrow. Hence  its experimental 
observation will  require resolution in $E_m$ on the order of a few MeV.}.

Our calculations show that, in general, 3N correlations are strongly affected by pair distortion 
and FSI effects. Pair distortion, having a qualitative features of 3N correlation, 
strongly affects  both types of 3N correlations, while the FSI contributes strongly in type 3N-I 
SRC and has a  diminished impact on type 3N-II SRCs only at parallel kinematics.

Finally, for near transverse kinematics starting at  $p_m\ge 400$~MeV/c,  the $A(e,e'N)X$ reactions 
represent an ideal tool for studying the structure of FSIs. This feature becomes especially 
valuable for large $Q^2>4$~GeV$^2$ for studies of color coherence phenomena for which we 
expect a decrease of  $T_S$ with an increase of $Q^2$ as opposed to the $Q^2$ independence of
$T_S$ in the GEA.

\begin{figure}[ht]
\vspace{-1.2cm}
\centerline{\epsfig{height=12cm,width=12cm,file=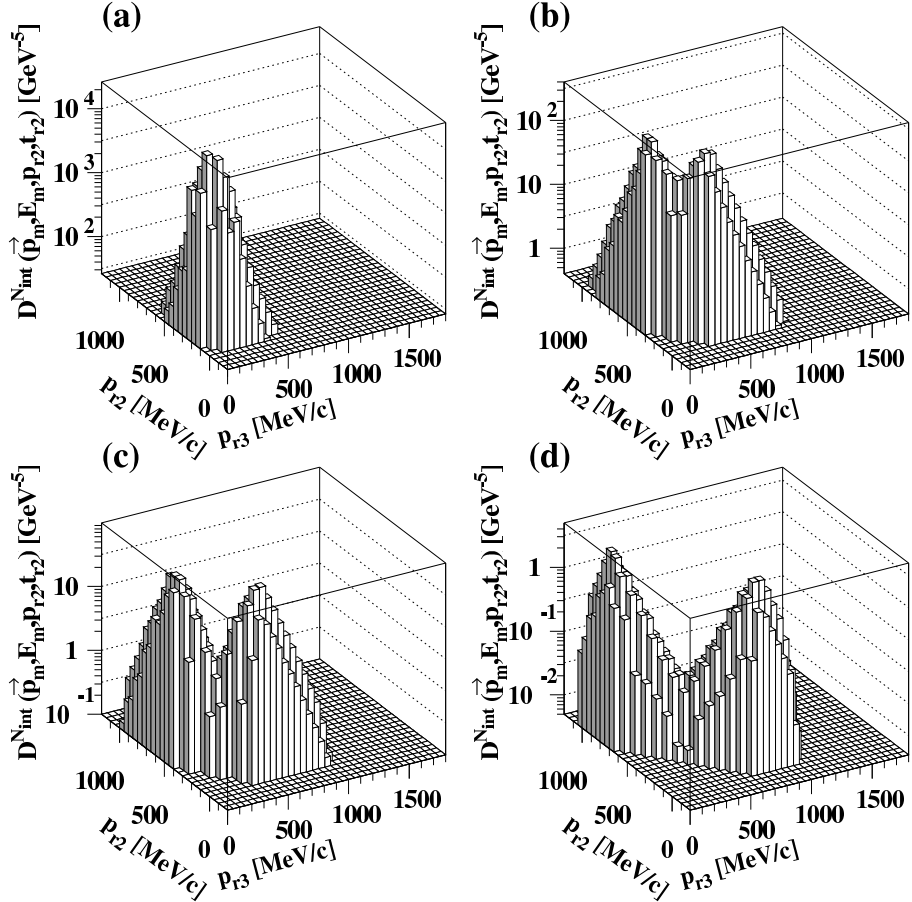}}
\caption{Dependence of the decay function on the momenta of the two recoil protons 
$p_{r2}$ and $p_{r3}$ at different cuts of missing momentum of the knocked-out neutron.
(a)$p_m>150$~MeV/c, (b)$p_m>300$~MeV/c, (c)$p_m> 400$~MeV/c and (d)$p_m > 700$~MeV/c.}
\label{decay_pwia_n}
\end{figure}

\subsection{Decay Function}
The decay function is practically an unexplored quantity and the experiments that will allow us 
to extract it from the exclusive cross section data are only emerging\cite{LW}.
Our main motivation in these numerical analyses  is to highlight those significant features 
of the decay function  that are related to the short-range structure of the ground state nuclear 
wave function as well was to the structure of reinteractions between the nucleons in the 
final state of the reaction.

We will consider the partially integrated decay function, which will allow us to remove 
the $\delta$-function in Eqs.(\ref{Dia}) and (\ref{Ddwia}). Namely, we consider,
\begin{equation}
D_{int}^N({\bf p_m},E_m,p_{r2},t_{r2}) = \int D^N({\bf p_m},E_m,{\bf p_{r2}},t_{r2})p^2_{r2}
d\Omega_{r2},
\label{Dfun-I}
\end{equation}
where  $D^N({\bf p_m},E_m,{\bf p_{r2}})$ is defined in 
Eqs.(\ref{Dia}) or (\ref{Ddwia}). This integration takes into account the fact that 
one of the components of ${\bf p_m}$ or ${\bf p_{r2}}$ is not independent and 
is fixed  from the energy conservation condition for the quasi-elastic disintegration of $^3He$. 
Therefore, $D_{int}$ represents a quantity which  could be extracted directly from a 
$^3He(e,e'NN)N$ experiment. Furthermore, we will refer to $D_{int}$ as a decay function.

\begin{figure}[ht]
\vspace{-0.6cm}
\centerline{\epsfig{height=12cm,width=12cm,file=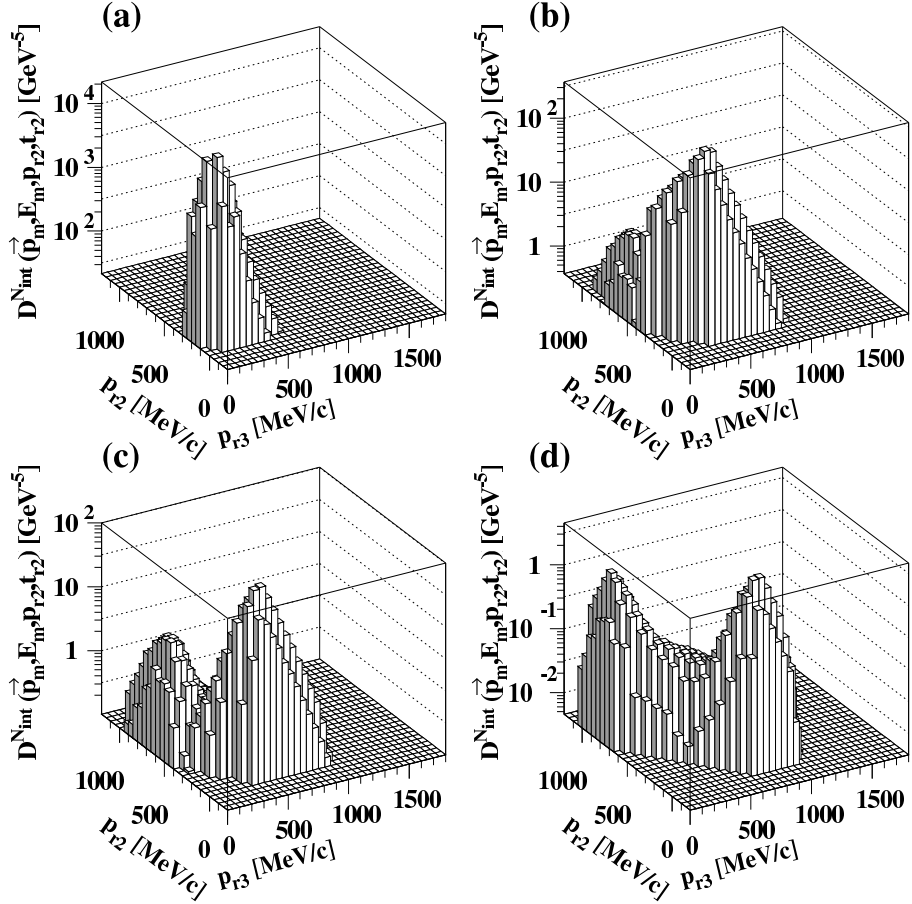}}
\caption{The same as in Fig.\ref{decay_pwia_n} for proton and neutron recoil 
with momenta  $p_{r2}$ and $p_{r3}$ respectively.} 
\label{decay_pwia_p}
\end{figure}

We start by analyzing the decay  function in the PWIA, focusing on those features 
that are related to  SRC signatures  of the ground state wave function which are described 
in Figs.\ref{2Nsrc} and \ref{3Nsrc}. As we observed in the previous section, 
type 2N-I SRCs exhibited  measurable  (though broad) correlation between 
$E_m$ and $p_m$ (starting at $p_m\ge 300$~MeV/c), with the peak of the $E_m$ distribution defined 
by Eq.(\ref{empm_2N-I}). These correlations could be understood qualitatively as a result of 
strong short range interactions between the struck nucleon (with momentum  $p_m$) and 
one of the recoil nucleons in the nucleus with the third nucleon having a relatively 
small momentum  in the mean field of the 2N SRC pair.  To check this picture, we calculate 
the decay function distribution 
as a function of $p_{r2}$ and $p_{r3}$ imposing different cuts on the missing momentum, 
$p_m$.

Figs.\ref{decay_pwia_n} and \ref{decay_pwia_p} present the results of the calculation of 
the decay function of the reactions corresponding to the knock-out of the neutron and proton 
respectively. 
The (a), (b), (c) and (d) parts correspond to the $p_m$ cuts at $p_m>150$, $300$, $400$ and 
$700$~MeV/c respectively. The calculations clearly show emerging type 2N-I correlations 
between the knocked-out nucleon and one of the spectator nucleons 
with an increase of $p_m$. These correlations are dominating the landscape of the 
($p_{r2}$, $p_{r3}$) 
distribution   once $p_m\ge 300 MeV/c$.  Fig.\ref{decay_pwia_p} shows also that
the $pn$ SRCs are significantly larger than the $pp$ correlations.  Thus, 
measuring the decay function will allow a direct check of the relative strength of $pp$ and $pn$
correlations. We conclude, from the discussions of Figs.\ref{decay_pwia_n} and 
\ref{decay_pwia_p},  that the decay function is strikingly more sensitive  to the 
SRC  than the $E_m-p_m$ correlation  observed in the spectral function.

\begin{figure}[ht]
\vspace{-0.8cm}
\centerline{\epsfig{height=10cm,width=10cm,file=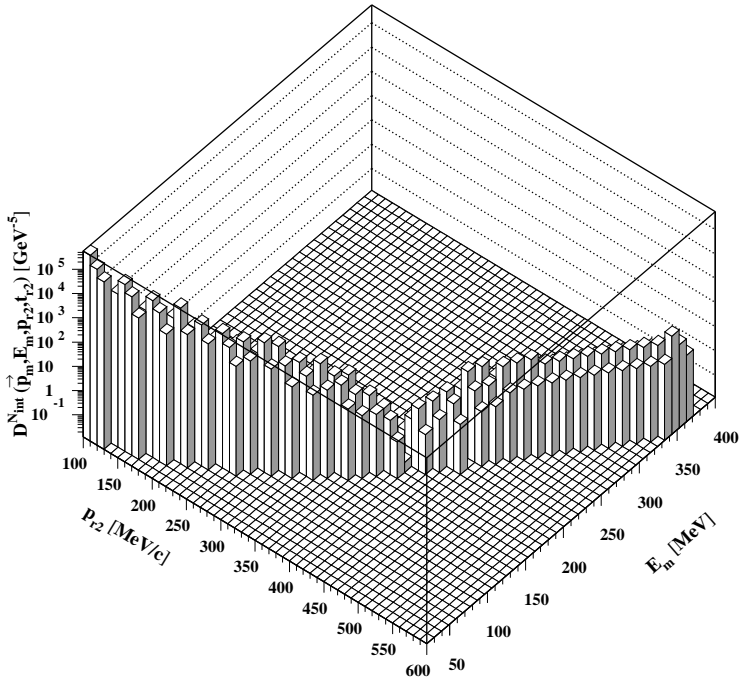}}
\caption{The ($E_m,p_{r2}$) distribution of the decay function for the neutron knockout 
reaction. The missing momentum is restricted to $p_m < 50$~MeV/c.}
\label{decay_2N-II}
\end{figure}

Next we analyze the features of the decay function related to the type 2N-II correlations. 
From the consideration of the spectral function in the previous section, we learned that 
reactions corresponding to the knock-out of the neutron are best suited for studies 
of type 2N-II correlations since, at small $p_m<100$~MeV/c and large $E_m$, the spectral 
function exhibits a  minimum associated with the node in the relative momentum distribution 
of the recoil $pp$ pair in the $S$-state.  In Fig.\ref{decay_2N-II} we show the 
($E_m,p_{r2}$) distribution of the decay function which demonstrates strong  
correlations between  the minimum in the $E_m$ distribution and the value of recoil nucleon 
momenta at small $p_m$ ($<50$~MeV/c in the calculation).

\begin{figure}[ht]
\vspace{-0.6cm}
\centerline{\epsfig{height=12cm,width=12cm,file=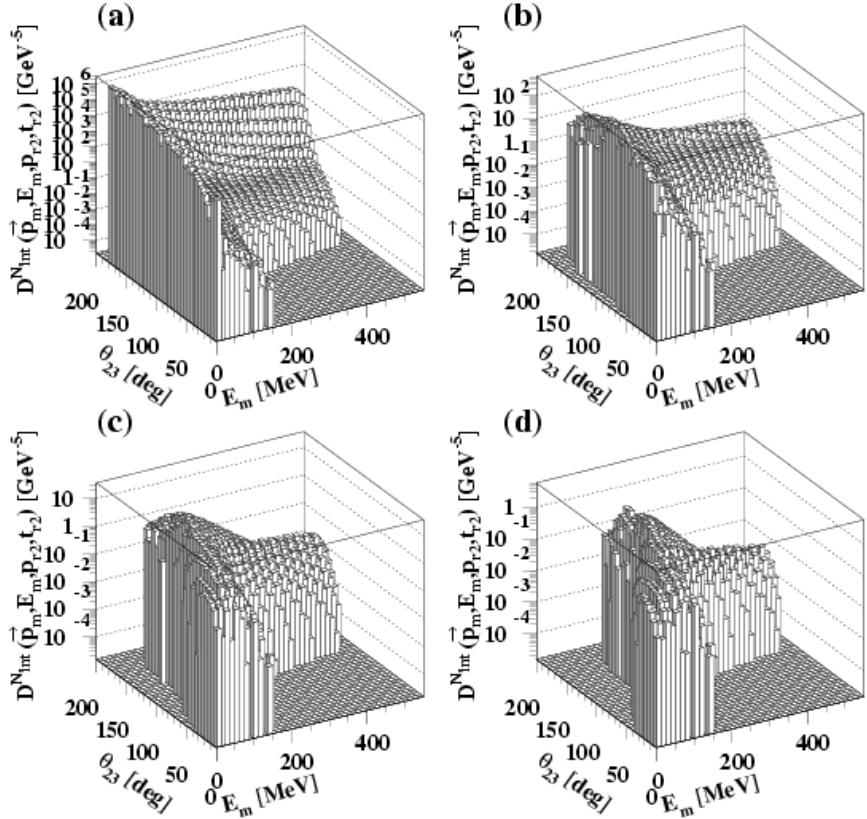}}
\caption{The dependence of the decay function of the reaction $^3He(e,e'pp)n$ on the relative angle 
between the  recoil $pn$ nucleons, $\theta_{23}$,  and  the missing energy, $E_m$, for different 
cuts of missing momentum, $p_m$. (a) -- no cuts on $p_m$, (b) -- $p_m > 300$~MeV/c, 
(c) -- $p_m > 500$~MeV/c,
(d) -- $p_m > 700$~MeV/c.} 
\label{decay_src}
\end{figure}

Turning now  to the three nucleon correlations, we observe that for both types of 3N 
SRCs (Fig.\ref{3Nsrc}) the two recoil nucleons have comparable momenta. This situation 
corresponds kinematically to the area around the saddle   in Figs.\ref{decay_pwia_n} and 
\ref{decay_pwia_p}.
One can  check whether the expectation that type 3N-I and 3N-II correlations will 
dominate in the dynamics of three nucleon correlations is justified. For this we observe that 
the configurations of Fig.\ref{3Nsrc} are characterized by a 
distinctive angular relations between the two recoil nucleons: for type 3N-I SRC 
two recoil nucleons 
will be produced at small relative angle (almost parallel) while for type 3N-II SRC the 
relative angle of the recoil nucleon momenta is $\approx 120^o$. 
In Fig.\ref{decay_src} we analyze the dependence of the decay function on the relative angle 
between the recoil nucleons, $\theta_{23}$,  and the missing energy, $E_m$, for different values of 
missing momentum, $p_m$.

\begin{figure}[ht]
\vspace{-0.8cm}
\centerline{\epsfig{height=11cm,width=11cm,file=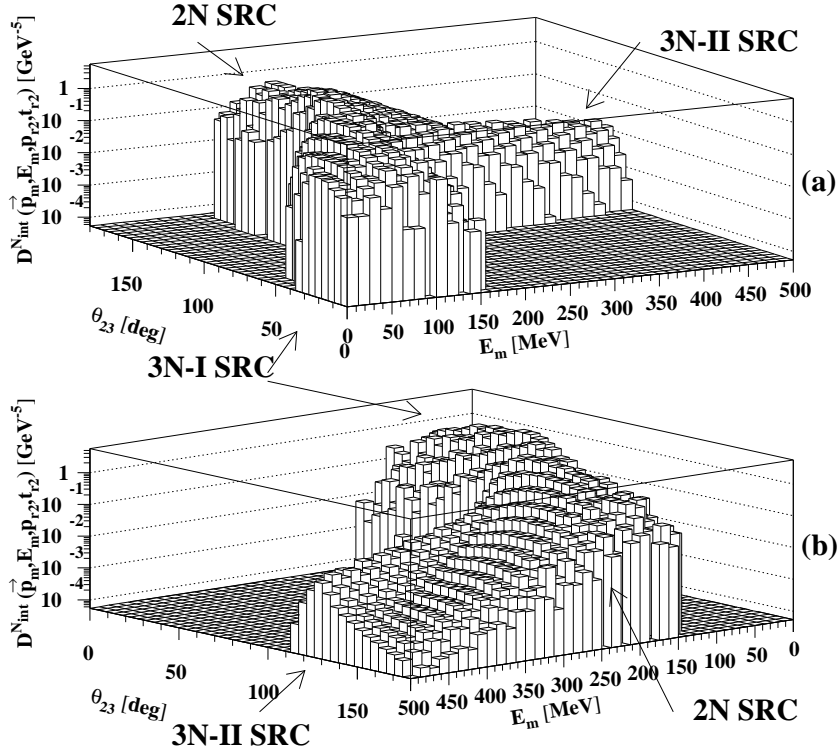}}
\caption{Reproduction of Fig.\ref{decay_src}d from  different viewpoints to 
emphasize the signatures of type 3N-I~(a) and type 3N-II~(b) correlations.}
\label{decay_src_ext}
\end{figure}

Fig.\ref{decay_src}(c) and (d) clearly show an emergence of peaks at small $E_m$ and 
$\theta_{23}$ ( type 3N-I SRCs) and at large $E_m$ and $\theta_{23}\approx 120^0$ ( type 3N-II 
SRCs). The appearance  of the peaks is more clearly seen in Fig.\ref{decay_src_ext} where 
Fig.\ref{decay_src}~(d) is demonstrated from  different viewpoints.  
Fig.\ref{decay_src_ext} shows also the peak at $\theta_{23}\approx 180^0$ and at moderate values of 
$E_m$ which corresponds to the type 2N-I correlation in which only one of the recoil nucleons 
is correlated with the knocked-out nucleon and is produced in the direction 
opposite to the direction of ${\bf p_m}$.

Moving to a more quantitative discussion of correlation properties of the decay function, 
in the following subsections we study the following question: 
To what degree do the detected properties of the recoil nucleon in coincidence with 
the knocked-out nucleon  reflect the properties of preexisting short range 
configurations in the ground state nuclear wave function? 

\subsubsection{Pair Distortion Effects}
\begin{figure}[ht]
\vspace{-0.6cm}
\centerline{\epsfig{height=12cm,width=12cm,file=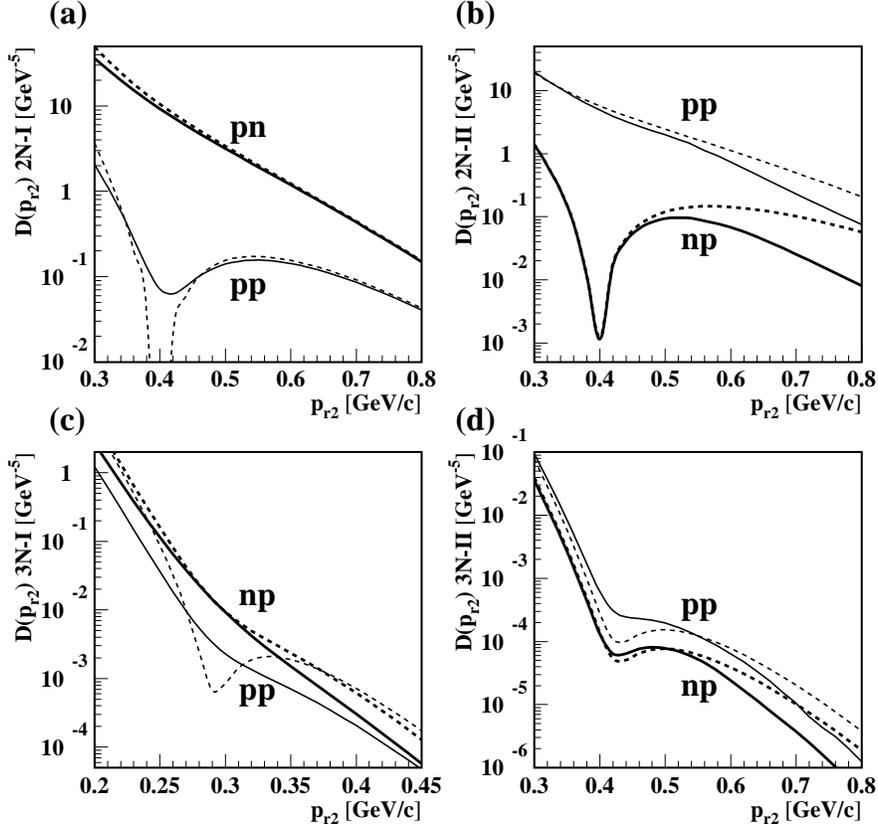}}
\caption{Dependence of the decay function on the momentum of the registered recoil nucleon, $p_{r2}$,
in $^3He(e,e'N_fN_{r2})N_{r3}$ reactions.(a), (b), (c) and (d) correspond to types 2N-I, 2N-II, 3N-I 
and 3N-II kinematics respectively. Dashed lines -- PWIA prediction, solid lines - IA predictions. 
Two pairs of curves in each figure correspond to different compositions of detected nucleons 
($N_f,N_{r2}$).}
\label{decay_srcs_pd}
\end{figure}

In Fig.\ref{decay_srcs_pd} we discuss the effects of the reinteraction between recoil 
nucleons (pair distortion) in the $p_{r2}$-momentum distribution of the decay function 
for all four types of correlations\footnote{Herewith, we will label calculations
by $(N_f,N_{r2})$, in which $N_f$ and $N_{r2}$ describe the type of the knocked-out 
and recoil nucleons with momenta $\bf p_{f}=\bf p_m+\bf q$ and $\bf p_{r2}$ respectively.}. 
In  Fig.\ref{decay_srcs_pd}(a) one 
observes a significant difference in the yields of the recoil proton or neutron for a  
reaction in which one of the protons is knocked out  from the type 2N-I correlations. 
Fig.\ref{decay_srcs_pd}(b) shows the momentum distribution of the recoil proton when proton  
(labeled ``pp'' in the figure) or neutron (labeled ``np'') is knocked-out in the kinematics of
type 2N-II correlations.

\noindent
This figure demonstrates  significant pair distortion effects as compared to 
the type 2N-I correlations.
Figs.\ref{decay_srcs_pd}(c) and (d) demonstrate the momentum distributions from type 3N-I and II 
SRC's respectively. As in the case of the spectral function, one observes that, in general,
the pair distortion interferes significantly with the three nucleon correlations.
It is worth noting, however, that, due to the depleted  interaction strength in the $pp$ channel 
at relative momenta $\sim 400$~MeV/c, pair distortion effects are suppressed in type 
3N-II correlations 
for the neutron knock-out reaction in a recoil momentum  range of $300-550$~MeV/c 
(see Fig.\ref{decay_srcs_pd}(d)). We will discuss in Sec.\ref{3NForce} how this observation 
could be used to explore the type 3N-II kinematics for investigation of three nucleon 
forces in $^3He$.

Also, Fig.\ref{decay_srcs_pd} reveals an additional feature that allows us to discriminate between  
2N and 3N SRC signals. This feature is the relative abundance of pp and pn pairs in different 
correlations.
For 2N correlations one observes significantly smaller yield associated with $pp$ correlations as 
compared to $pn$ correlations while for 3N correlations these yields are comparable. 

\medskip
\medskip

\subsubsection{Final State Interaction}

The inclusion of the final state interaction of the knocked-out nucleon with residual nucleons 
removes the isotropy of the decay function with respect to the momentum vector of the 
virtual photon, ${\bf q}$. Staying in the framework of consideration of  type 2N and 
3N correlations, we investigate FSI effects in angular and momentum dependences of the decay 
function for each type of correlation. As in Sec.\ref{Sec.Spec.FSI}, we consider 
the kinematics with fixed $Q^2=4$~GeV$^2$. 

\medskip
\medskip

\noindent
{\bf A. Type 2N-I Correlations:}

\vspace{0.2cm}

\begin{figure}[ht]
\vspace{-0.8cm}
\centerline{\epsfig{height=12cm,width=12cm,file=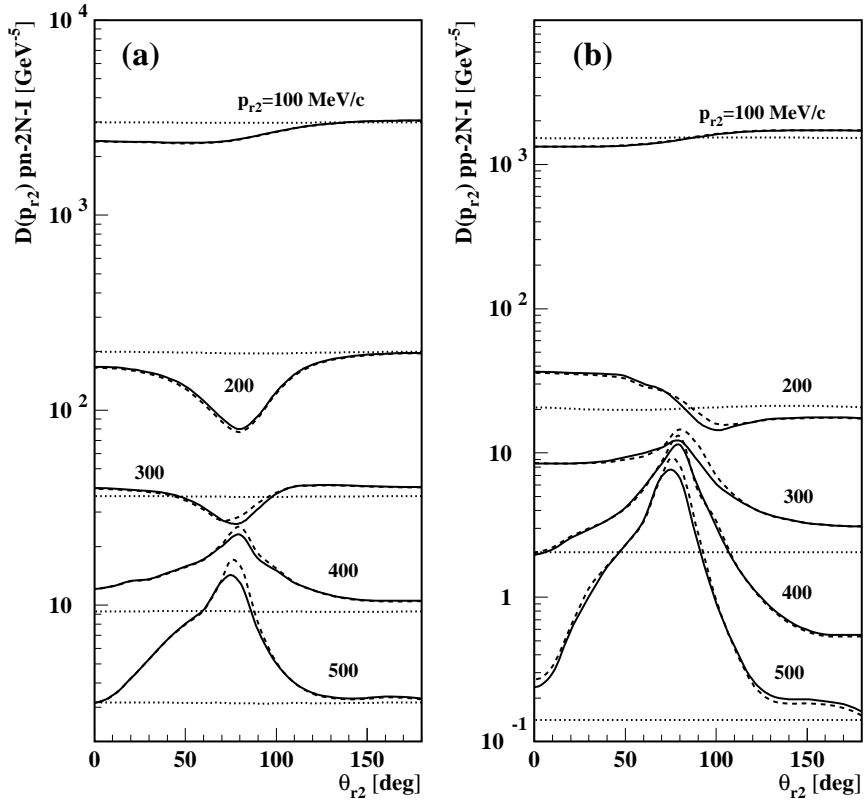}}
\caption{Dependence of the  decay function on the angle of the detected recoil nucleon, 
$\theta_{r2}$, with respect to $\bf q$ for different and fixed values of $p_{r2}$.
The decay function is calculated for type 2N-I SRC kinematics. (a) -- $(N_f,N_{r2})\equiv (p,n)$, 
(b)-- $(N_f,N_{r2})\equiv (p,p)$. Dotted lines - IA, dashed lines - IA+FSI1~(Single Rescattering only), 
solid line - IA+FSI. Note that for figure (b) the IA contribution at $p_{r2}=400$~MeV/c is not 
shown since it is negligible due to the node in the $pp$ momentum distribution. 
}
\label{decay_2n-i_fsi_th}
\end{figure}

For the type 2N-I correlations, we consider first the dependence of the decay function on the 
production angle of recoil nucleon $N_{r2}$ with respect to the direction of the 
virtual momentum ${\bf q}$. Fig.\ref{decay_2n-i_fsi_th} shows these
dependences for different values of recoil nucleon (neutron~(a) and proton~(b)) momentum 
$p_{r2}$ for  the reaction with knocked-out proton. Calculations 
clearly show the transition of FSI from the screening regime at momenta $p_{r2}\le 300$~MeV/c 
to the double scattering regime at $p_{r_2}\ge 400$~ MeV/c.  This picture is clearly 
reminiscent  the double  scattering signatures of the 
deuteron electro-disintegration\cite{deuteron,ggraphs,treview}. 

Calculations also predict very different angular dependences for neutron (a) and proton (b) 
production in recoil kinematics. It is easy to check that the maximal FSI 
effects are predicted at $\alpha_{r2}={E_{r2}-p^z_{r2}\over m}=1$ which  is 
analogous to the  maximums of the FSI for the spectral functions at 
$\alpha_m=1$ observed in Sec.\ref{Sec.Spec.FSI}.
Fig.\ref{decay_2n-i_fsi_th} shows also that our best chances to extract  genuine information 
about type 2N-I SRC, is to concentrate on anti-parallel ($\theta_{r2}=0^0$) and parallel 
($\theta_{r2}=180^0$) kinematics. It is worth noting that the range of $\theta_{r2}$ where
FSI, are small is broader for the $\theta_{r2}=180^0$ case.

\begin{figure}[ht]
\vspace{-0.6cm}
\centerline{\epsfig{height=11cm,width=11cm,file=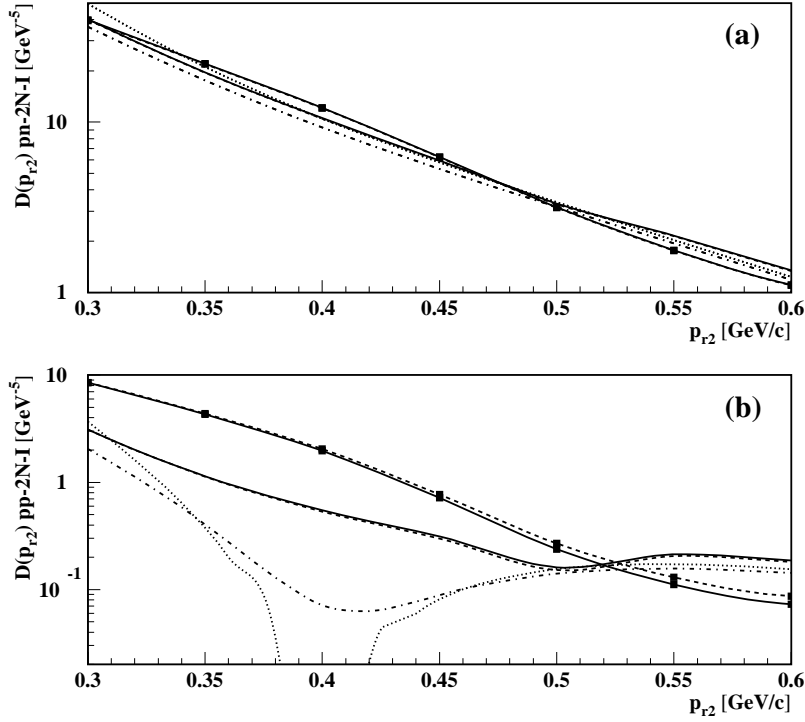}}
\caption{Dependence of the decay function on the momentum of the recoil nucleon, $p_{r2}$ 
calculated for the kinematics of type 2N-I correlations (a) -- $(N_f,N_{r2})\equiv (p,n)$ 
and (b)-- $(N_f,N_{r2})\equiv (p,p)$. Doted lines - IA prediction, dashed and solid lines 
for IA+FSI1 and IA+FSI predictions respectively for parallel $\theta_m=0^0$ kinematics. 
Curves with square labels correspond to IA+FSI1 and IA+FSI predictions for 
anti-parallel $\theta_m=180^0$ kinematics.}
\label{decay_2n-i_fsi_pr2}
\end{figure}

Fig.\ref{decay_2n-i_fsi_pr2} represents the $p_{r2}$ momentum dependences of the decay function
for $\theta_{r2}=180^0$  and $\theta_{r2}=0^0$ (marked curves) kinematics. Calculations 
predict qualitatively different momentum distributions for correlated $pn$ (a) and $pp$ (b) 
pairs. While for $pn$  at  both $\theta_{r2}=180^0$ and  $\theta_{r2}=0^0$ one observes 
qualitatively similar momentum distributions, for $pp$ pairs they are significantly different. 
This difference can be understood from Eq.(\ref{pZ}), which defines the effective 
longitudinal component of missing momentum entering in the rescattering amplitude. 
For type 2N-I kinematics it can be written 
as $p_Z = -p_{r2}cos(\theta_{r2}) + {q_0\over q}(E_m+{p_{r2}^2\over 4m})$.  At 
$\theta_{r2}=180^0$ ($\theta_{r2}=0^0$), $p_Z > (<) p^z_{r2}$ and the FSI term is defined by the 
effective momentum which is larger (smaller) than measured momentum, $p_{r2}$. 
As a result, the FSI is suppressed at $\theta_{r2}=180^0$ (backward) kinematics as 
compared to the $\theta_{r2}=0^0$ (forward) kinematics. For the $pp$ pair this difference is 
very dramatic due to the node in the relative momentum distribution of the 
$pp$ pair.

It is worth noting that no significant triple scattering (double rescattering) 
contribution is observed in Figs.\ref{decay_2n-i_fsi_th} and \ref{decay_2n-i_fsi_pr2}. 
This serves as an additional indication that in type 2N-I kinematics the reaction is 
defined predominantly   by two-body interactions, thus representing a nearly 
ideal framework for studies of  the 2N processes both in terms of the short range 
correlations and single rescattering  processes.

\medskip

\noindent
{\bf B. Type 2N-II Correlations:}

\vspace{0.2cm}

\begin{figure}[ht]
\vspace{-0.8cm}
\centerline{\epsfig{height=11cm,width=11cm,file=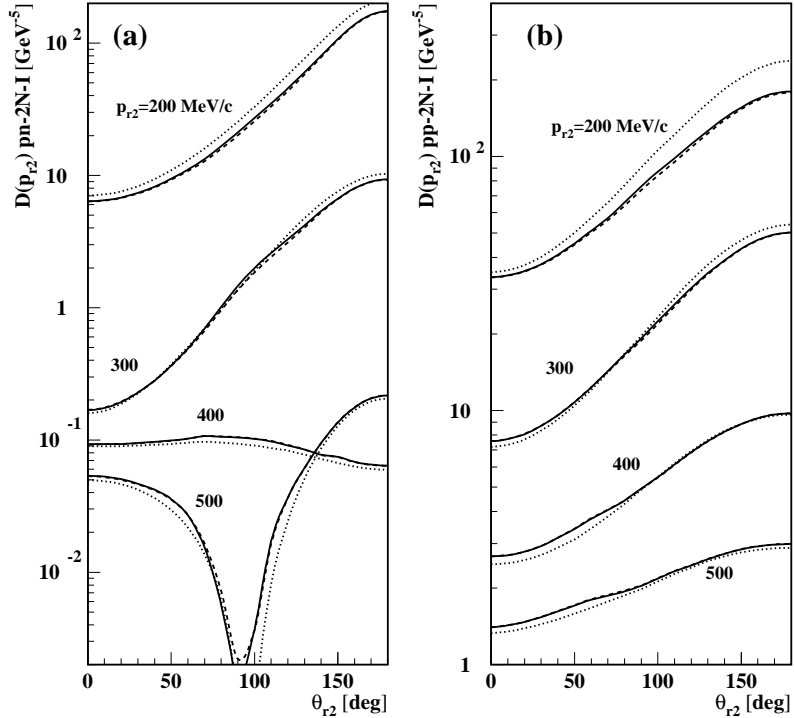}}
\caption{Dependence of the  decay function on the angle of the detected recoil nucleon, 
$\theta_{r2}$ with respect to $\bf q$ for different and fixed values of $p_{r2}$.  
The decay function is calculated for type 2N-II SRC kinematics with missing momentum fixed at 
$p_m=100$~MeV/c and $\theta_m=0^0$. 
(a) -- $(N_f,N_{r2})\equiv (n,p)$, 
(b)-- $(N_f,N_{r2})\equiv (p,p)$. Dotted lines - IA, dashed lines - IA+FSI1, 
solid line - IA+FSI.}
\label{decay_2n-ii_fsi_th}
\end{figure}

\begin{figure}[ht]
\vspace{-0.8cm}
\centerline{\epsfig{height=10cm,width=12cm,file=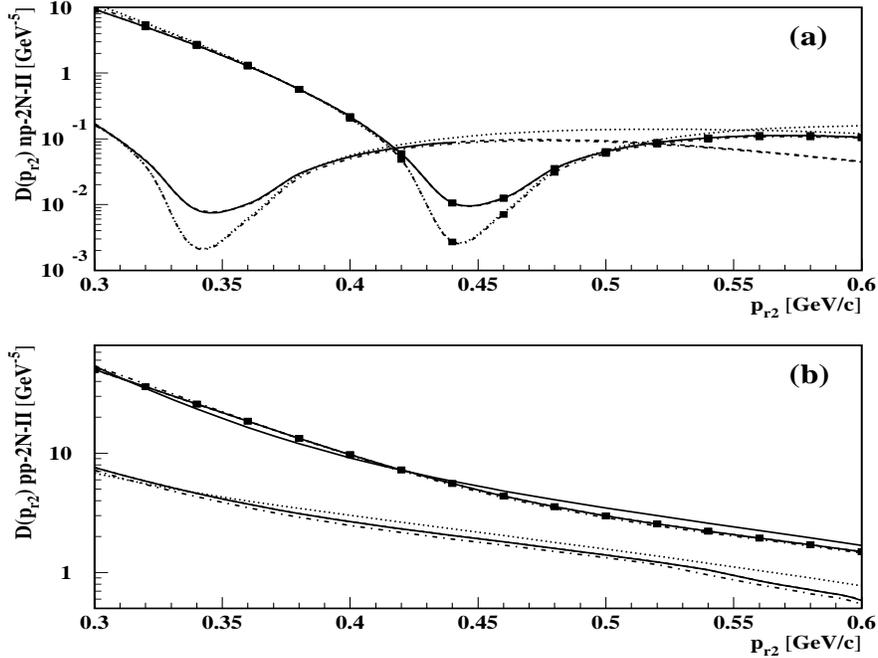}}
\caption{Dependence of the decay function on the momentum of recoil nucleon, $p_{r2}$ 
calculated for the kinematics of type 2N-II correlations with missing momentum fixed at 
$p_m=100$~MeV/c and $\theta_m=0^0$. (a) -- $(N_f,N_{r2})\equiv (n,p)$ 
and (b)-- $(N_f,N_{r2})\equiv (p,p)$. Doted lines - IA prediction, dashed and solid lines 
for IA+FSI1 and IA+FSI predictions respectively for $\theta_{r2}=0^0$ kinematics. 
Curves with square labels corresponds to IA+FSI1 and IA+FSI predictions for 
anti-parallel $\theta_{r2}=180^0$ kinematics.}
\label{decay_2n-ii_fsi_pr2}
\end{figure}

To isolate type  2N-II correlations we identify kinematics similar to one presented 
in Fig.\ref{decay_srcs_pd}(b), in which missing momentum is fixed to $p_m=100$~MeV/c in 
${\bf q}$ direction. In Fig.\ref{decay_2n-ii_fsi_th} we consider the $\theta_{r2}$ dependence
of the decay function for both $pp$~(a) and $pn$~(b) recoil pairs.  Since, in these 
kinematics, the knocked-out nucleon is on average  at a larger distance from the recoil 2N pair, 
one expects the lesser FSI effects due to rescattering of the knocked-out nucleon off 
the recoil nucleons. 
Calculations presented in Fig.\ref{decay_2n-ii_fsi_th} confirm this expectation. The same pattern 
can be seen in the momentum distribution plot in Fig.\ref{decay_2n-ii_fsi_pr2} which 
shows that the FSI is diminished  practically for the whole region of recoil 
nucleon momenta, $p_{r2}$, of interest.

Comparison of type 2N-I and 2N-II correlation  cases demonstrate that type 2N-II makes 
the best case for probing the node of the relative momentum distribution  in the $pp$ 
pair in the $^3He$ wave function (see Figs.\ref{decay_2n-ii_fsi_th}(a) and 
\ref{decay_2n-ii_fsi_pr2}(b)). However, a definitive answer on whether the node can be 
observed in the experiment requires a careful accounting of non-pole effects in the pair-distortion 
contribution. This consideration is out of scope of present paper and a dedicated study of the 
node effects will be presented elsewhere\cite{node}.

As in the case of type 2N-I correlations, the effects of double rescattering are marginal which 
indicates again the  feasibility of isolating two-body effects without complication,  due to 
double rescattering of knocked-out nucleon off both recoil nucleons. 

\medskip
\medskip

\noindent
{\bf C. Type 3N-I Correlations:}

\vspace{0.2cm}

\begin{figure}[ht]
\vspace{-0.8cm}
\centerline{\epsfig{height=11cm,width=11cm,file=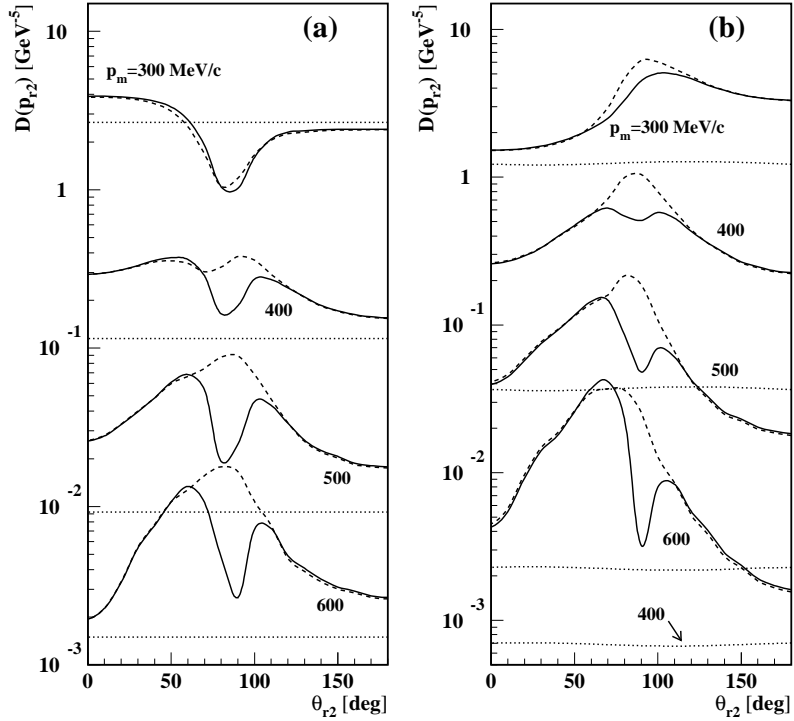}}
\caption{Dependence of the  decay function on the angle of the detected recoil nucleon, 
$\theta_{r2}$, with respect to $\bf q$ 
for different and fixed values of $p_{m}$. 
The decay function is calculated for type 3N-I SRC kinematics such that the recoil momentum 
of the detected nucleon is opposite to the missing momentum direction and defined as 
$p_{r2} = {p_m\over 2} + 50$~MeV/c. (a) -- $(N_f,N_{r2})\equiv (n,p)$, 
(b)-- $(N_f,N_{r2})\equiv (p,p)$. Dotted lines - IA, dashed lines - IA+FSI1, solid line - IA+FSI.
}
\label{decay_3n-i_fsi_th}
\end{figure}

A consideration of the angular dependence of the decay function for type 3N-I correlations
in Fig.\ref{decay_3n-i_fsi_th}
reveals a significant  effect of FSIs for almost all angles of 
pair production (or $\theta_m$). The FSI dominates especially at transverse kinematics 
at $\alpha_m=1$ which reveals a significant contribution from the double rescattering 
processes starting at $p_m\ge 400$~MeV/c. Production of $pp$ pairs (Fig.\ref{decay_3n-i_fsi_th}(a)) 
in parallel kinematics 
provides the best condition for probing type 3N-I SRCs (albeit not without considerable pair 
distortion effects) starting at $p_m\ge 600$~MeV/c. This can be understood qualitatively.
Since each rescattering transfers $k_z\approx \Delta, k_t\le 500$~MeV/c momentum 
(predominantly in transverse direction; $k_t>\Delta$), and the rescattering  amplitude falls 
exponentially with $k^2_t$, it is kinematically infeasible to rescatter two nucleons above 
600~MeV/c in the backward direction.

\begin{figure}[ht]
\vspace{-0.6cm}
\centerline{\epsfig{height=10cm,width=12cm,file=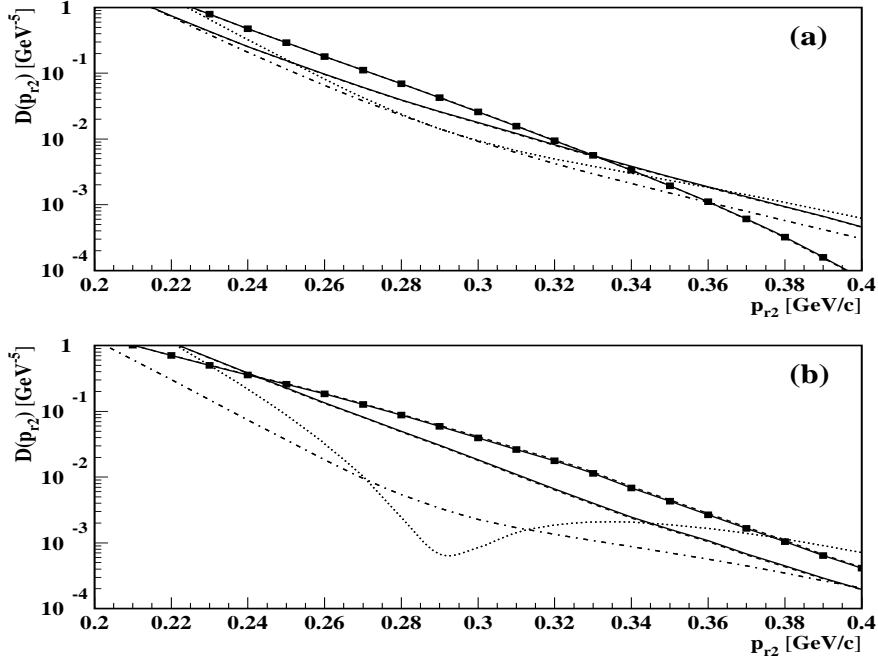}}
\caption{Dependence of the decay function on the momentum of recoil nucleon, $p_{r2}$, 
calculated for the kinematics of type 3N-I correlations. The relation between missing 
and recoil momenta are the same as in Fig.\ref{decay_3n-i_fsi_th}.   (a) -- $(N_f,N_{r2})\equiv (n,p)$ 
and (b)-- $(N_f,N_{r2})\equiv (p,p)$. Dotted lines - IA prediction, dashed and solid lines 
for IA+FSI1 and IA+FSI predictions respectively for $\theta_{r2}=180^0$ kinematics. 
Curves with square labels corresponds to IA+FSI1 and IA+FSI predictions for 
anti-parallel $\theta_{r2}=0^0$ kinematics.}
\label{decay_3n-i_fsi_pr2}
\end{figure}

The effects discussed in the previous paragraph
can be seen in more detail in the momentum distribution in Fig.\ref{decay_3n-i_fsi_pr2} which 
confirms again that the only reasonable chance to extract the original momentum 
distribution in type 3N-I correlation 
is to measure the coherent recoil $pp$ pair production in
the parallel kinematics $\theta_{m}=0^0$. 
Note that the significant contribution from double rescattering processes at 
transverse kinematics can be also understood qualitatively. To produce two coherent 
nucleons at large angle it is more preferable for the knocked-out nucleon 
to scatter consecutively  off spectator nucleons. It contributes maximally in 
the transverse kinematics due to the eikonal nature  of NN rescattering which 
dominates at $\alpha_m=1$. This situation is reminiscent of the 
dynamics relevant to the form-factors of few-body systems measured in high momentum 
transfer reactions, in which case one needs significant rescatterings between constituents 
of the system to produce coherent combination of the subsystem.

\medskip
\medskip

\noindent
{\bf D. Type 3N-II Correlations:}

\vspace{0.2cm}

\begin{figure}[ht]
\vspace{-0.8cm}
\centerline{\epsfig{height=11cm,width=12cm,file=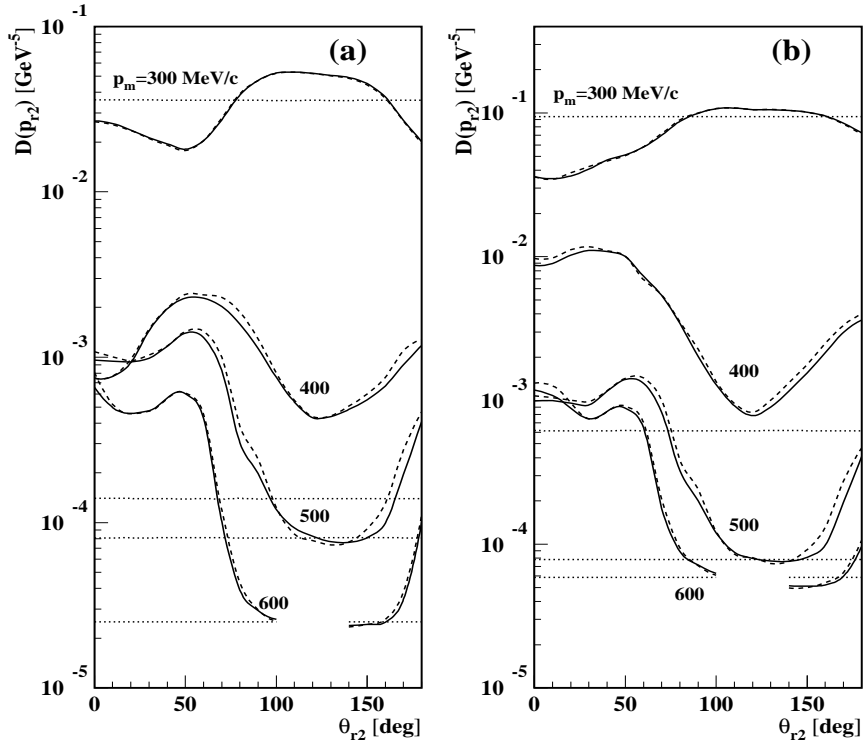}}
\caption{Dependence of the  decay function on the angle of the detected recoil nucleon, 
$\theta_{r2}$, 
with respect to $\bf q$ for different and fixed values of $p_{r2}$. 
The decay function is calculated for type 3N-II SRC kinematics such that $p_{m}=p_{r2}=p_{r3}$ and 
relative angles between these momentum vectors are $120^0$. Here, $\theta_{r2}=120^0$ corresponds 
to $\theta_m=0^0$. (a) -- $(N_f,N_{r2})\equiv (n,p)$, 
(b)-- $(N_f,N_{r2})\equiv (p,p)$. Dotted lines - IA, dashed lines - IA+FSI1, solid line - IA+FSI.}
\label{decay_3n-ii_fsi_th}
\end{figure}

In considering the angular dependence of the decay function for  type 3N-II correlations 
in Fig.\ref{decay_3n-ii_fsi_th}, we
observe overall  large  FSI effects  except in the kinematics in which both 
recoil nucleons are produced in the backward direction compared to that of $\bf q$. 
In this  figure this corresponds to the situation when the recoil nucleon 
with momentum $p_{r2}$ is produced at $\theta_{r2}=120^0$ and the fast 
nucleon is knocked out in parallel kinematics ($\theta_m=0^0$). This automatically 
puts the  production angle of the second recoil nucleon at $\theta_{r3}=120^0$ in 
the other half of the scattering plane.

One can understand the suppression of FSI in this kinematics qualitatively. It is 
very improbable with one single rescattering to produce two nucleons in the backward 
hemisphere of knocked-out energetic nucleon. One may expect that double rescattering 
can contribute to production of such configurations. However, as it will be 
discussed in the next section, it is dominant only 
at $\alpha_{r2}\approx \alpha_{r3}\approx 1$ which is significantly away from the 
considered kinematics. 
Note that the different angular dependence for recoil $pp$~(a) and $pn$~(b) pairs at
momentum range $300-500$~MeV/c is related to the qualitative difference in the relative 
momentum distribution of these pairs (namely to an existing node in $pp$ distribution). 
Note that recoil nucleon angles  $110 < \theta_{r2} <   130$ at $\theta_m=0^0$ and $p_{r2}=600$~MeV/c 
are kinematically forbidden since in this case $\alpha_m+\alpha_{r2}+\alpha_{r3}\ge 3$. 
However, the $600$~MeV/c curve shows that the FSI is small at broader angular ranges starting at 
$80^0$ to $150^0$. This may be very important for probing larger internal momenta in type 
3N-II SRCs without substantial FSI effects.

\begin{figure}[ht]
\vspace{-0.6cm}
\centerline{\epsfig{height=12cm,width=12cm,file=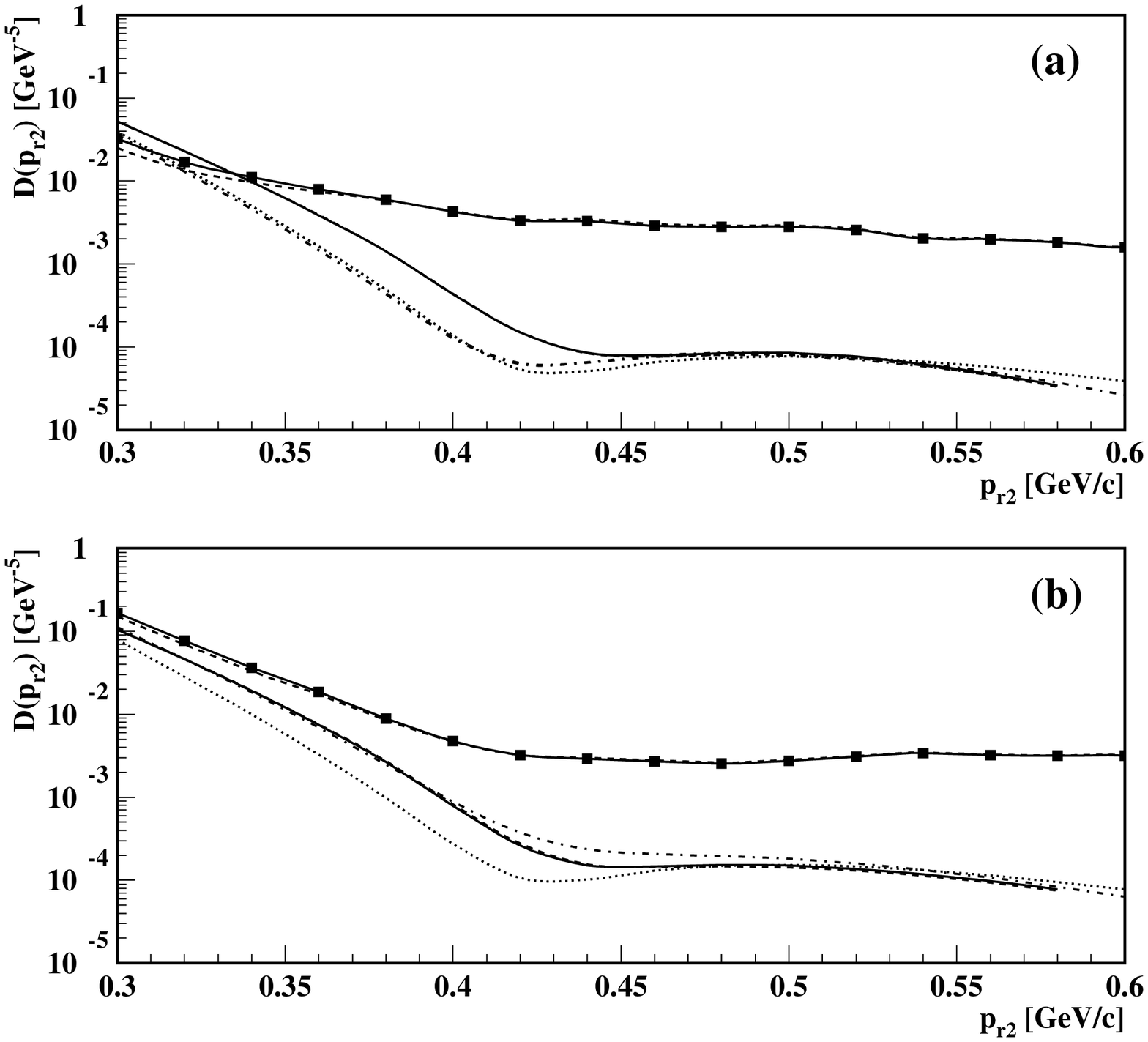}}
\caption{Dependence of the decay function on the momentum of the recoil nucleon, $p_{r2}$, 
calculated for the kinematics of type 3N-II correlations. The relation between missing 
and recoil momenta is the same as in Fig.\ref{decay_3n-ii_fsi_th}.   
(a) -- $(N_f,N_{r2})\equiv (n,p)$ 
and (b)-- $(N_f,N_{r2})\equiv (p,p)$. Doted lines - IA prediction, dashed and solid lines 
for IA+FSI1 and IA+FSI predictions respectively for $\theta_m=0^0$ ($\theta_{r2}=120^0$) kinematics. 
Curves with square labels correspond to IA+FSI1 and IA+FSI predictions for 
anti-parallel $\theta_m=180^0$~($\theta_{r2}=60^0$) kinematics.}
\label{decay_3n-ii_fsi_pr2}
\end{figure}

The momentum distribution of the decay function in Fig.\ref{decay_3n-ii_fsi_pr2}
confirms the observed, in Fig.\ref{decay_3n-ii_fsi_th}, smallness of FSIs at 
parallel kinematics with two recoil nucleons produced in the backward hemisphere at 
120$^0$.  This situation provides a unique window for accessing type 3N-II correlations. 
They will be discussed  in more details in Sec.\ref{3NForce}.

In  Fig.\ref{decay_3n-ii_fsi_pr2} we also compare the momentum distribution for anti-parallel 
($\theta_m=180^0$) kinematics when recoil nucleons are produced in forward $60^0$ 
angles. Here we observe  significant FSIs which enhances the cross section by 
almost an order of magnitude.

\subsubsection{Double Rescattering Effects}
\label{Sec.DFSI}

Previous considerations demonstrated that in practically all cases the maximal 
FSI is generated in kinematics in which $\alpha_m=1$. To enhance the effects of 
double rescattering, relative to the single rescattering contribution, the strategy is to 
keep $\alpha_m=1$ and choose kinematics  in which two recoil nucleons are produced 
in symmetric and transverse kinematics.  Such  configurations boost the double 
rescattering effects, since in these cases the most economic way to produce two 
recoil nucleons with large transverse momentum 
is to have two consecutive rescatterings of the knocked-out nucleon off 
the spectator nucleons in $^3He$.    
One such kinematics corresponds to the type 3N-I correlations, in which 
two recoil nucleons are produced with  almost  vanishing relative 
momentum. Fig.\ref{decay_3n-i_fsi_th} demonstrates a large  contribution to double 
rescattering when two coherent recoil nucleons are produced at almost transverse angles 
with respect to the direction of $\bf q$.

\begin{figure}[ht]
\vspace{-0.6cm}
\centerline{\epsfig{height=10cm,width=12cm,file=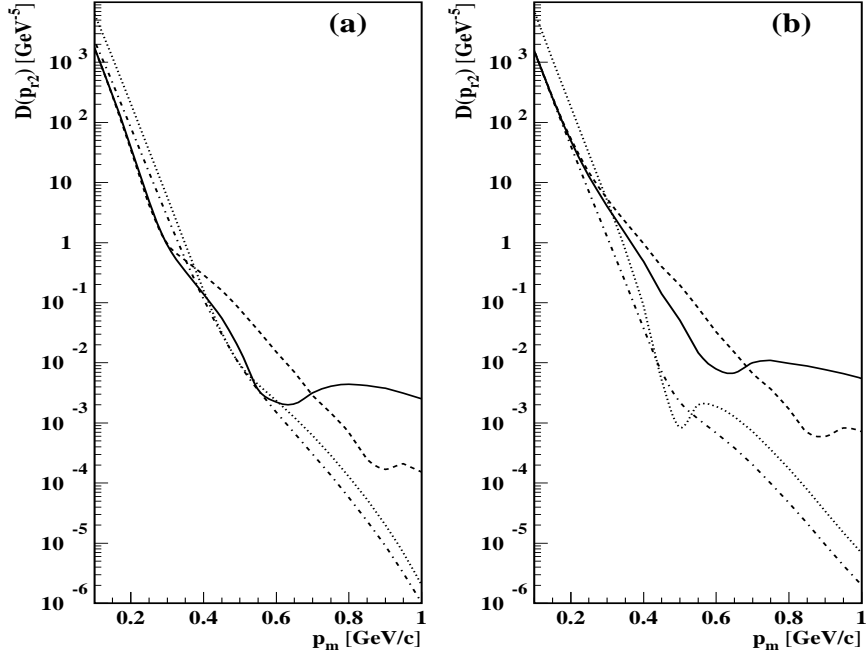}}
\caption{Dependence of the decay function on $p_m$ at $\alpha_m=1$ for type 3N-I kinematics.
The direction of the recoil nucleons momenta is opposite to that of the missing momentum direction 
with $p_{r2} = {p_m\over 2}+50$~MeV/c.  (a) -- $(N_f,N_{r2})\equiv (n,p)$ 
and (b)-- $(N_f,N_{r2})\equiv (p,p)$. Dotted line -- PWIA, dash-doted -- IA, dashed -- IA+FSI1 and 
solid -- IA+FSI predictions.}  
\label{decay_n-nn_double}
\end{figure}

In Fig.\ref{decay_n-nn_double}  we consider further 
the kinematics of the  type 3N-I correlation by calculating  
the $p_m$ dependence of the decay function at $\alpha_m=1$. The result is the significant 
enhancement of the double rescattering effects starting at $p_m\ge 300$~MeV/c. In the range of 
$300 \le p_m \le 700$~MeV/c, double rescattering screens the single rescattering contribution 
through its destructive  interference with the single rescattering amplitude 
(see Eqs.(21) and (27) in Ref.\cite{SASF1}).  
However, starting with $p_m \ge 700$~MeV/c, our calculations show that the decay function is 
determined predominantly by the square of the double rescattering amplitude 
of Eq.(27) in Ref.\cite{SASF1}. 
Since the internal nucleon momenta in the $^3He$ wave function are small in the double rescattering 
amplitude, the relativistic effects are expected to be small in spite of $p_m>700$~MeV/c.

\begin{figure}[ht]
\vspace{-0.8cm}
\centerline{\epsfig{height=10cm,width=12cm,file=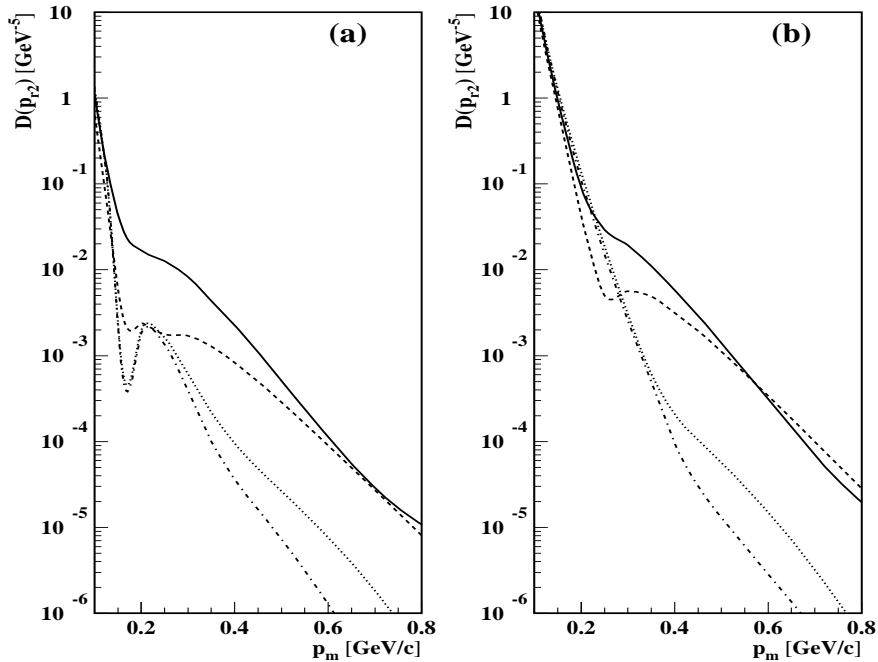}}
\caption{Dependence of the decay function on $p_m$ at $\alpha_m=\alpha_{r2}=\alpha_{r3}=1$.
(a) -- $(N_f,N_{r2})\equiv (n,p)$ and (b)-- $(N_f,N_{r2})\equiv (p,p)$. 
Dotted line -- PWIA, dash-doted -- IA, dashed -- IA+FSI1 and  solid -- IA+FSI predictions.}
\label{decay_n-n-n_double}
\end{figure}

Another  kinematical region which provides the symmetric configuration for recoil nucleons is the one
close to the type 3N-II correlations in which the light cone momentum fractions of all three 
nucleons are chosen $\alpha_m=\alpha_{r2}= \alpha_{r3}=1$, and two recoil nucleons are produced in 
the opposite halves  of the scattering plane. In this case, one again expects the enhanced 
contribution from double rescattering. The analysis of single (Eq.(21) in Ref.\cite{SASF1}) and 
double rescattering (Eq.(27) in Ref.\cite{SASF1}) amplitudes shows that the double rescattering 
contribution is maximized in  the kinematics where the interference of 
single and IA~(Eq.(11) in Ref.\cite{SASF1}) amplitudes 
cancel the square of the single rescattering amplitudes, which happens at $p_m\approx 300$~MeV/c.
This can be seen in 
Fig.\ref{decay_n-n-n_double} which displays missing momentum dependence of the decay function in 
the above kinematical region discussed above. One observes that at $p_m\approx$~300 MeV/c 
the double rescattering diagram is dominant. It is interesting to note that in this case  the 
dominant part comes from the interference between IA (Eq.(11) in Ref.\cite{SASF1}) and 
double~(Eq.(27) in Ref.\cite{SASF1}) rescattering amplitudes.\\

\noindent {\bf Digression:  Color Transparency:} The ability to isolate the double rescattering 
contribution in $^3He$ electro-disintegration may play a significant role in the ongoing 
studies of color transparency (CT) phenomena.  The main premise of CT is that at sufficiently 
large $Q^2$ the knocked-out nucleon 
is produced in a point like configuration~(PLC) which, due to the color neutrality of 
the object, will have a reduced hadronic interaction strength. Thus, CT phenomena will be 
manifest in the decreasing of the $f_{PLC-N}$ amplitude of rescattering  with an increase of $Q^2$. 
This is in contrast to the $Q^2$ independence of $f_{NN}$ in  eikonal approximation.

Presently, two complementary  experimental approaches 
are used to  find the signatures for CT phenomenon. One is the attenuation experiments\cite{Ent} 
in which nuclear transparency is measured in $(e,e'N)$ reaction off nuclei with $A\ge 2$ and the 
other is the deuteron electro-disintegration reactions\cite{KE} which are aimed at 
measuring the single rescattering terms in $d(e,e'N)N$  reactions. While attenuation experiments 
are sensitive to $\sim f_{PLC-N}$, the single-rescattering experiments can provide sensitivity 
up to  $\sim f_{PLC-N}^2$. The possibility of isolating the double rescattering term in $^3He$ 
electro-disintegration will allow us to gain unprecedented sensitivity - up to $\sim f_{PLC,N}^4$ 
(see Eq.(27) in Ref.\cite{SASF1}). Furthermore,
isolating the double rescattering amplitude will allow us  
to address such an intricate questions as whether the first rescattering  will destroy the 
PLC coherence formed  by the initial high $Q^2$ $\gamma^*N$ scattering.

\subsubsection{Probing Three Nucleon Forces}
\label{3NForce}
Three nucleon forces~(3NFs) are one of the most elusive features in nuclear physics. 
The existence of them for the triton was assumed for the first time by Wigner\cite{Wigner}  
even before the triton was discovered experimentally. There is  little theoretical guidance for 
systematic building of 3NFs and the main experimental evidence used to constraint the 
different 3NF models is the binding energy of $A=3$ nuclei (for review of the present status of 
3NFs see \cite{Friar}). 
The importance of 3NFs  was emphasized in the studies of the binding energies of $A=3$ 
nuclei.
Furthermore, it allowed for improvements in the calculation of binding energies 
of nuclei $A = 4 -9$\cite{PPWC}. The 3NFs can  significantly modify the present models of 
equations of state of nuclear matter\cite{HeisPand},  
therefore, understanding 3NFs can have a significant impact on our understanding of the physics 
relevant to the superdense nuclear matter such as neutron stars. 
However, as it is  mentioned in Ref.\cite{Friar}, an accuracy of $1\%$ in calculations are 
needed in order to systematically disentangle 3NFs forces from the overwhelming 2N interactions 
in few nucleon  systems.

In our consideration of 3NFs, we proceed from the assumption that, just by the nature of 3N forces, 
they will dominate in type 3N-II correlations. 
Conversely, in the case of the 
type 2N-I correlations, 
it is possible to suppress 3NFs significantly by restricting the momentum of the third spectator 
nucleon to be close to the zero.
Thus, our expectation is that  one should observe significantly different contributions from 3NFs in 
3N-II and 2N-I correlations.   
Furthermore, we recall our discussions of type 2N-I and 3N-II correlations in Secs.V.B(2A) and 
V.B(2D) where we found that one can significantly suppress FSI effects in type 2N-I correlations 
in parallel kinematics $(\theta_m=0^0)$ (see Fig.\ref{decay_2n-i_fsi_pr2}) and in type 3N-II 
correlations in kinematics where we choose two recoil nucleons to be produced in the backward 
hemisphere with respect to ${\bf q}$ (see Fig.\ref{decay_3n-ii_fsi_pr2}).

\begin{figure}[ht]
\vspace{-0.8cm}
\centerline{\epsfig{height=10cm,width=12cm,file=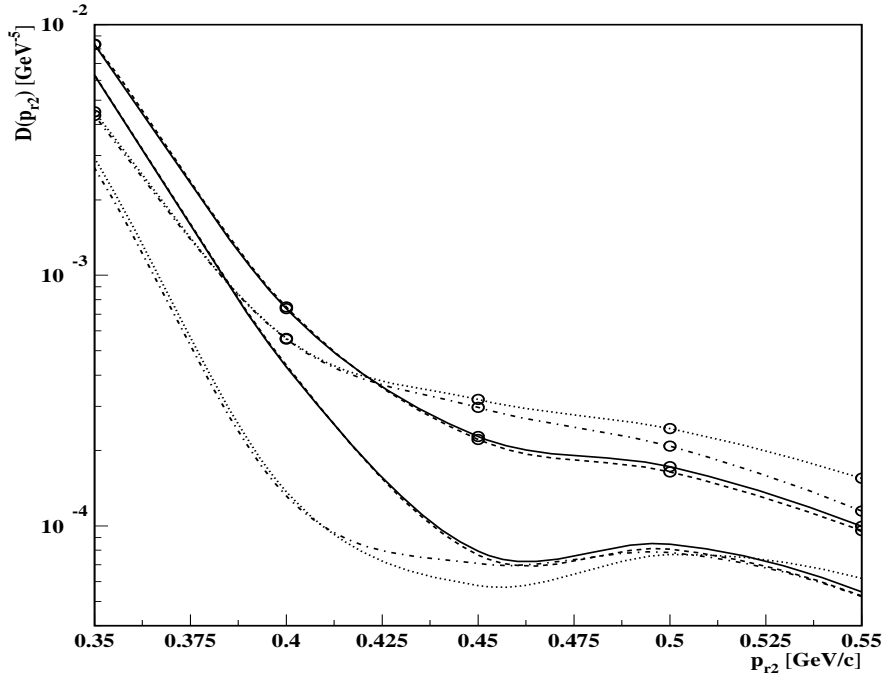}}
\caption{Dependence of the decay function on the momentum of the recoil nucleon, $p_{r2}$,
calculated for the kinematics of type 3N-II correlations and $(N_f,N_{r2})\equiv (n,p)$. 
The relation between missing and recoil momenta is the same as in Fig.\ref{decay_3n-ii_fsi_th}, 
and the $r2$ recoil proton is detected at $\theta_{r2}=120^0$  while the angle of missing 
momentum $\theta_m=0^0$.
Dotted line -- PWIA, dash-doted -- IA, dashed -- IA+FSI1 and  solid -- IA+FSI predictions. 
Curves with  square labels correspond to the same contributions 
with three-nucleon forces included in the calculation of the $^3He$ wave function.}
\label{decay_3n-ii_3nf}
\end{figure}

Thus we expect that these kinematic windows are optimal for 
probing 3NFs. To quantify our statement, in Fig.\ref{decay_3n-ii_3nf} we compare
the momentum distributions of the decay function calculated in the type 3N-II kinematics 
when two recoil protons are produced at $120^0$  relative to the $\bf q$. 
Calculations are done for two cases: in one case we have only $NN$ interactions 
while in the other case 3NFs are included according to the Tucson-Melbourne model~\cite{TM}.
Our calculations show a factor of two difference between calculations including 
only 2N forces and calculation including 2N+3NF forces.

\begin{figure}[ht]
\vspace{-0.6cm}
\centerline{\epsfig{height=10cm,width=12cm,file=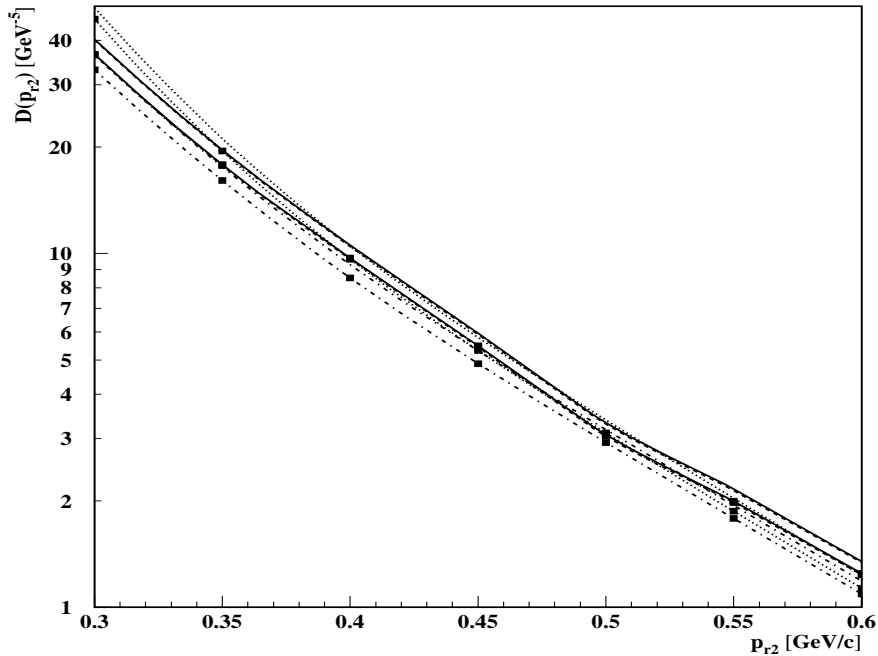}}
\caption{Dependence of the decay function on the momentum of the recoil nucleon,
$p_{r2}$, for type 2N-I kinematics at $\theta_{r2}=180^0$~($\theta_m=0^0$). 
$(N_f,N_{r2})\equiv (n,p)$. Dotted line -- PWIA, dash-doted -- IA, dashed -- IA+FSI1 
and  solid -- IA+FSI predictions. Curves with  circular labels correspond to the same contributions 
with three-nucleon forces included in the calculation of the $^3He$ wave function.}
\label{decay_2n-i_3nf}
\end{figure}

In contrast, the calculations in the kinematics of the type 2N-I correlations presented in 
Fig.\ref{decay_2n-i_3nf}, 
using the  same two models of NN interactions, predict little difference between momentum 
dependences of the decay function.

These considerations show that we can identify the domains   in the kinematics of  type 3N-II 
correlations where  FSI effects are relatively small and one has a good chance to extract the 
genuine information about 3NFs. This assumes one is doing 
simultaneous measurement also in  2N-I kinematics where the same 3NFs will 
give a negligible contribution.

\section{Summary}
\label{V}
Based on generalized eikonal approximation, developed in the part-I of this work\cite{SASF1},
for high $Q^2$ electro-disintegration of $A=3$ system we perform  numerical studies of 
exclusive $^3He(e,e'N)NN$ reactions.

As an input, in our numerical studies  we use: (i)~The Bochum group's~\cite{Bochum}
calculation of ground state $^3He$ wave function for different sets of realistic NN interaction 
potentials as well as calculations which explicitly include three nucleon forces in the 
ground state. (ii) The SAID group's\cite{SAID} parameterization 
of low-to-intermediate energy NN scattering amplitudes to calculate the two-nucleon 
continuum state wave function which is needed in order to evaluate the  interaction between 
recoil nucleons in the final state of the reaction (pair distortion effects). 
(iii)~We use our updated parameterization of  high energy NN scattering amplitudes in the  
calculation of the small angle rescattering of struck nucleon off spectator nucleons\cite{tabra}.

To describe the  exclusive $^3He(e,e'N)NN$ reaction we use the formalism of the decay function  
which is related to the conventional spectral function through the integration of the phase 
space of the recoil nucleons.

In the numerical analyses of both spectral and decay functions, we concentrate on studies of 
two main types of  2N and 3N correlations. 
For 2N correlations, we consider the ones in which the struck nucleon is initially  
correlated with one of the recoil nucleons while the third nucleon is spatially isolated 
(type 2N-I SRC). Also for the case of 2N correlations we consider  
the case in which the two recoil nucleons are in 2N correlation with the struck nucleon in the 
mean field of the SRC pair (type 2N-II SRC). For 3N correlations we consider the correlations 
in which struck nucleon with large missing momenta is correlated with the pair of coherent nucleons 
(type 3N-I SRC). For the 3N correlations we also consider the case
in which all three nucleons have relative momenta exceeding the characteristic 
mean field momentum in the nucleus and have a relative angle $\approx 120^0$ (type 3N-II SRC). 

In discussing the spectral function, we demonstrate that it exhibits several unique features related 
to the structure of 2N correlations. These are the correlation between missing momenta and missing 
energy and the minimum in the spectral function associated with the node in the 
relative momentum distribution of the $pp$ pair. These results are in agreement with  previous  
analyzes of spectral function (see e.g. \cite{Ciofi}). There are, however, no clear cut signatures 
in the spectral 
function related to the 3N correlations. Within PWIA, 3N SRCs reveal only through 
the strength of spectral function at very high values of missing momenta and/or energy.
Our new result in considering the spectral function is that the strength related to the 
3N SRCs is practically washed out by the pair distortion and the FSI effects. 

Considering the decay function, which is practically an unexplored quantity, we observe that within 
PWIA it clearly exhibits  the main features of 2N and 3N correlations. Subsequent analysis of pair 
distortion and FSI effects revealed that the additional degrees of freedom associated with 
the full detection of the decay products of the reaction allows us to pinpoint unambiguously
the kinematics in which the FSI or SRC plays a dominant role.  Our conclusion is that 
the comprehensive  study of the decay function in exclusive reactions will allow 
an unprecedented access to the 2N and 3N correlations in the nuclei.

We highlight two particular cases. One is the possibility to isolate  double rescattering effects  
which can provide us  a new and powerful tool in studying  color transparency phenomena.  
The other is identifying a kinematic window that will allow us to probe directly the effects 
associated with three-nucleon forces in the ground state wave function  of $^3He$.

Our overall conclusion is that the investigation of the decay function opens up a completely new 
venue in studies of short range nuclear properties and allows us to discriminate between 
different orders of final state reinteractions. The latter will provide  a powerful tool in studies 
of color transparency phenomena.

\begin{acknowledgements}
We are grateful to Andreas Nogga for providing us with Bochum group's wave 
functions of $^3He$. Special thanks to  Richard Arndt and Igor Strakovsky for their help 
in using the SAID program. We thank Ted Rogers for careful reading of the manuscript and 
for many valuable comments.
This work is supported  by  DOE grants under contract DE-FG02-01ER-41172 and DE-FG02-93ER40771 as 
well as by the Israel-USA Binational Science Foundation Grant.
M.M.S. gratefully acknowledges also a contract from Jefferson Lab under which this work was 
done. The Thomas Jefferson National Accelerator Facility (Jefferson Lab) is operated 
by the Southeastern Universities Research Association (SURA) under DOE contract 
DE-AC05-84ER40150.
\end{acknowledgements}

 \end{document}